\documentclass[prd,12pt,nofootinbib,preprint]{revtex4}
\usepackage[T1]{fontenc}
\usepackage{amsmath,amssymb}
\usepackage{epsfig}
\usepackage{url}
\usepackage{graphicx}
\usepackage[usenames,dvipsnames]{color}
\usepackage{slashed}
\usepackage[colorlinks,citecolor=blue]{hyperref}
\usepackage{color}
\usepackage{cancel}

\def\a {\alpha}
\def\b {\beta}

\def\l {\lambda}
\def\L {\Lambda}

\def\bar {\overline}

\def\be {\begin{equation}}
\def\ee {\end{equation}}
\def\beq {\begin{equation}}
\def\eeq {\end{equation}}
\def\bea {\begin{eqnarray}}
\def\eea {\end{eqnarray}}

\newcommand{\besub}{\begin{subequations}}
\newcommand{\eesub}{\end{subequations}}

\def\beq{\begin{equation}}
\def\eeq{\end{equation}}
\def\barr{\begin{array}}
\def\earr{\end{array}}

\begin{document}
\preprint{HRI-RECAPP-2018-001, NCTS-PH/1802}
\title{Ameliorating the Higgs mass fine-tuning problem with multi-Higgs doublet models}
\author{Nabarun Chakrabarty}
\email{nchakrabarty@cts.nthu.edu.tw, chakrabartynabarun@gmail.com}
\affiliation{Regional Centre for Accelerator-based Particle Physics, 
Harish-Chandra Research Institute, HBNI, Chhatnag Road, Jhunsi, Allahabad - 211019, India} 

\affiliation{Physics Division, National Center for Theoretical Sciences, Hsinchu, Taiwan 30013, R.O.C}
\author{Indrani Chakraborty}
\email{indranichakraborty@hri.res.in}
\affiliation{Regional Centre for Accelerator-based Particle Physics, 
Harish-Chandra Research Institute, HBNI, Chhatnag Road, Jhunsi, Allahabad - 211019, India} 
\begin{abstract}
With no conclusive signal till date of the minimal supersymmetric and extra
dimensional models at the Large Hadron Collider (LHC), the issue
of fine-tuning of the Higgs mass still calls for some attention. It could be
very possible that the observed Higgs boson of mass 125 GeV has its origin 
in a non-supersymmetric extended scalar sector, and, it still has properties
strikingly similar to the Standard Model Higgs. In such cases, however, one
relies upon possible cancellations in the quadratic divergence of the Higgs mass to uphold naturalness. In this work, we have investigated this possibility in context of some two Higgs doublet and three Higgs doublet scenarios. 
\end{abstract} 
\maketitle
\section{Introduction}
\label{intro}
The performance of the Standard Model (SM) in describing the fundamental interactions up to a certain scale being excellent, it is still believed to be an effective theory above which new physics (NP) takes over. Besides the other reasons like non-existence of suitable dark matter candidate, vanishing neutrino mass, Baryon asymmetry etc. the "Fine-tuning problem" of Higgs mass is one of the major contributors behind this belief.

Absence of any particular symmetry protecting the scalar masses like the gauge symmetry and the chiral symmetry, that protect the masses of the gauge bosons and fermions respectively, leads to quadratically-divergent self-energy corrections to the masses of the scalars. One of the conventional approaches to yield a finite scalar mass out of a large radiative correction is to engineer a huge cancellation between the bare mass term and the radiative correction term, {\em i.e.} to {\em "fine-tune''} these two terms accordingly by the virtue of some known symmetry. The approach adopted by us to ensure the finite mass of the scalar would be in the reverse way. We shall assume that by virtue of some yet-to-be discovered symmetry, it is possible to make the coefficient of the quadratically divergent correction exactly zero or negligible. This condition was named after Veltman who first proposed it, and is termed as "Veltman condition" \cite{Veltman:1980mj}.
The SM scalar potential
\bea
V(\Phi) = - \mu^2 \Phi^\dag \Phi + \lambda (\Phi^\dag \Phi)^2 \,
\eea
leads to the quadratically divergent correction to the Higgs self-energy at one-loop as 
\bea
\delta m_h^2 
= \frac{\Lambda^2}{16 \pi^2} 
\left(6 \lambda + \frac {3}{4} g_1^2 + \frac{9}{ 4} g_2^2 - 6 g_t^2\right) \,.
\label{VC_SM_1}
\eea
where $g_1$, $g_2$ are the $U(1)_Y$ and $SU(2)_L$ gauge couplings and $g_t$ is the top quark Yukawa coupling. $\lambda$ and $\Lambda$ are Higgs self-coupling and cutoff scale respectively. The coefficient of $\frac{\Lambda^2}{16 \pi^2}$ in Eq.(\ref{VC_SM_1}) is termed as Veltman coefficient. The strict implementation of Veltman condition for the Higgs boson in SM leads to 
\bea
VC_h = \left(6 \lambda + \frac 3 4 g_1^2 + \frac 9 4 g_2^2 - 6 g_t^2\right) = 0 \,.
\label{VC_SM}
\eea
Thus the Veltman condition for the Higgs boson is far from being satisfied in SM,
because it needs Higgs mass $m_h \approx 316$ GeV. If one fixes the couplings 
in Eq.(\ref{VC_SM}) at electroweak (EW) scale, owing to the large negative contribution from top loop, the Veltman coefficient $VC_h$ becomes negative instead of zero. This  shortcoming of SM can be circumvented by extending the scalar sector of SM with additional singlet(s) \cite{Kundu:1994bs,Chakraborty:2012rb,Karahan:2014ola}, with more than one doublet \cite{Chakraborty:2014oma,Biswas:2016bth}, or with triplets \cite{Chakraborty:2014xqa}. There are plenty of studies available which employ this bottom-up approach to yield the finite scalar mass \cite{Antipin:2013exa,Masina:2013wja,Drozd:2012is,Pivovarov:2007dj}. Besides taking care of the quadratic divergence of the Higgs boson, one should also keep in mind that there would also be quadratic-divergences arising from the other scalars constituting the scalar spectrum of the model, therefore one should also try to pin down those quadratic-divergences in a similar way like the Higgs boson.

In this paper we aim to address the fine-tuning problem with two famous extensions of SM, namely, the two Higgs doublet models (2HDMs) \cite{Branco:2011iw,Bhattacharyya:2015nca} and the three Higgs doublet models (3HDMs) \cite{Felipe:2013ie,Machado:2010uc,Ivanov:2014doa,Ivanov:2012fp,Degee:2012sk}. We have considered four different types of 2HDM without tree-level flavour changing neutral currents (FCNCs), {\em i.e.} type-I , type-II , lepton specific and flipped 2HDMs. Besides we have also considered  $S_3$-symmetric 2HDM \cite{Das:2017zrm} and inert doublet models \cite{Branco:2011iw} for the sake of completeness of discussion. Likewise $A_4$-symmetric \cite{Felipe:2013ie,Machado:2010uc,Degee:2012sk}, $S_4$-symmetric \cite{Felipe:2013ie, Degee:2012sk} and $S_3$-symmetric 3HDMs \cite{Chakrabarty:2015kmt, Das:2014fea} have been considered as variants of 3HDMs. In a nutshell our objectives are the following :
\begin{itemize}
\item Searching for the parameter spaces of the models compatible with the theoretical and experimental constraints, which will be illustrated in detail in different sections.
\item To identify the lightest neutral Higgs boson of each model with 
the 125 GeV SM Higgs boson. Our primary concern is to cancel the quadratic divergence of the SM Higgs boson by satisfying the Veltman condition to an appropriate degree.
\item Once the Veltman condition for the SM Higgs boson is satisfied, with the same set of model parameters we try to satisfy the same for the other physical scalars present. If it is not possible, we try to fine-tune the tree-level mass term and the one-loop correction term accordingly, so that the 
quadratic-divergences of the other physical scalars remain under control.
\end{itemize}
The paper is structured as follows. In Section \ref{2HDM} we discuss all the model features including various constraints imposed on the parameter space of different types of 2HDM and the degree of compatibility with the Veltman conditions for various physical scalars in respective models. Section \ref{3HDM} deals with the same for various types of 3HDMs. Finally we summarize and conclude in Section \ref{Summary}.

\section{Two Higgs doublet models}
\label{2HDM}
The 2HDM \cite{Branco:2011iw,Bhattacharyya:2015nca} being one of the important extensions of Standard Model (SM), contains an extra $SU(2)$ doublet $\Phi_2$ having hypercharge $Y = +1$, apart from the SM Higgs doublet $\Phi_1$. The scalar spectrum of 2HDM consists of five physical scalars, {\em i.e.} the light Higgs $h$, the heavy Higgs $H$, the pseudoscalar Higgs $A$ and two charged Higgs $H^{\pm}$. If the scalar potential of 2HDM is $CP$-conserving, then one can assign definite $CP$-properties to the physical scalars, which makes the mass eigenstates identical with the $CP$-eigenstates. A generic 2HDM scalar potential \cite{Branco:2011iw,Bhattacharyya:2015nca} suffers from tree level FCNCs, which is alleviated using Glashow-Weinberg-Paschos (GWP) theorem \cite{Glashow:1976nt,Paschos:1976ay}. According to the theorem, quarks of given charges receive their masses by coupling with any one of the two doublets. This can be implemented by imposing $\mathbb{Z}_2$-symmetries on the scalars and fermions. Depending on the discrete symmetries imposed, there can be four types of 2HDMs as follows :
\begin{enumerate}
\item Type-I, for which all fermions couple with $\Phi_2$ and none with $\Phi_1$ ;
\item Type-II, for which up-type quarks couple to $\Phi_2$ , down-type quarks and charged leptons couple to $\Phi_1$; 
\item Type Y (sometimes called Type-III or Flipped), for which up-type quarks and charged leptons couple to $\Phi_2$ and down-type quarks couple to $\Phi_1$; 
\item Type X (sometimes called Type-IV or Lepton-specific), for which all charged leptons couple to $\Phi_1$ and all quarks couple to $\Phi_2$ ;
\end{enumerate}
In the subsections that follow, we discuss the different variants of the 2HDM
taken in this study.
\subsection{2HDM with $CP$-conserving potential}
\subsubsection{Scalar potential}
The most general renormalizable scalar potential of 2HDM involving two Higgs doublets $\Phi_1$ and $\Phi_2$ with hypercharge $Y = +1$ can be written as \cite{Branco:2011iw,Bhattacharyya:2015nca},
\bea
 V &=&  m_{11}^2 \Phi_1^\dag\Phi_1 + m_{22}^2 \Phi_2^\dag\Phi_2 - m_{12}^2 \left(\Phi_1^\dag\Phi_2 
  + \Phi_2^\dag\Phi_1\right)\nonumber\\
  && + \frac12\lambda_1\left(\Phi_1^\dag\Phi_1\right)^2+\frac12\lambda_2\left(\Phi_2^\dag\Phi_2\right)^2
     + \lambda_3 \left(\Phi_1^\dag\Phi_1\right)\left(\Phi_2^\dag\Phi_2\right) \nonumber\\
&&   + \lambda_4 \left(\Phi_1^\dag\Phi_2\right)\left(\Phi_2^\dag\Phi_1\right)+
\frac12\lambda_5\left[\left(\Phi_1^\dag\Phi_2\right)^2 + \left(\Phi_2^\dag\Phi_1\right)^2\right] \nonumber \\
&& \lambda_6 \left(\Phi_1^\dag \Phi_1\right) \left[\Phi_1^\dag \Phi_2 + \Phi_2^\dag \Phi_1\right] + \lambda_7 \left(\Phi_2^\dag \Phi_2\right) \left[\Phi_1^\dag \Phi_2 + \Phi_2^\dag \Phi_1\right]\,.
\label{2HDM_pot}
\eea
Where the doublets are parametrised as,
\bea
\Phi_i = \frac{1}{\sqrt{2}} \begin{pmatrix}
\sqrt{2} w_i^+ \\
v_i + h_i + i z_i
\end{pmatrix},~ {\rm for} ~ i = 1,2.
\eea
Here $m_{11}, m_{22}, \lambda_1, \lambda_2, \lambda_3, \lambda_4$ are real and $m_{12}, \lambda_5, \lambda_6, \lambda_7$ can in principle be complex. With real $m_{12}, \lambda_5$ and $\lambda_6, \lambda_7 = 0$, one can arrive at a $CP$-conserving potential \cite{Branco:2011iw,Bhattacharyya:2015nca}. We shall consider the  $CP$-conserving potential of 2HDM throughout our analysis. 

In this model, $\tan\beta$ is an important parameter and is defined by the ratio of the vacuum expectation value (VEV) of two doublets, {\em i.e.} $\tan\beta = \frac{v_2}{v_1}$. The couplings $\lambda_i$'s can be expressed in terms of the masses of the physical scalars, {\em i.e.} $m_h, m_H, m_A, m_{H^\pm}$ and the mixing angles $\alpha$ , $\beta$ \cite{Barroso:2013awa,Chakraborty:2015raa,Chakrabarty:2014aya}.

\subsubsection{Constraints imposed}
\begin{enumerate}
\item  {\textbf{\em{Stability conditions :}}} \\
The requirement that the 2HDM potential must be bounded from below along any field direction, leads to the following stability conditions connecting the quartic couplings in the scalar potential \cite{Branco:2011iw,Bhattacharyya:2015nca}.
\besub
\bea
\lambda_1, \lambda_2 \geq 0,  \\
\lambda_3 \geq  - \sqrt{\lambda_1 \lambda_2},  \\
\lambda_3 + \lambda_4 - |\lambda_5 | \geq - \sqrt{\lambda_1 \lambda_2} \,.
\eea
\eesub
\item  {\textbf{\em{Perturbativity and unitarity :}}}  \\
The upper limit on the quartic couplings ($|\lambda_i| \leq 4\pi ( i = 1,2,...5$)) and Yukawa couplings ($|y_i| \leq \sqrt{4\pi} ( i = t, b, \tau)$), makes the theory perturbative at the given scale \cite{Chakrabarty:2014aya}. The upper limits on the couplings are more constrained from unitarity requirements of the theory as can be found in \cite{Chakrabarty:2014aya}. 
\item {\textbf{\em{Oblique parameters and flavour constraints :}}} \\
The additional contribution of 2HDM to $S, T, U $ oblique parameters ($\Delta S , \Delta T , \Delta U$) arises due to the presence of the additional scalars of the model in loops. Among all these deviations, $\Delta T$ is most constrained and controls the mass splitting between the charged Higgs and the
neutral Heavy Higgs \cite{Bhattacharyya:2015nca,Chakrabarty:2014aya}. We have used $\Delta T = 0.09 \pm 0.13$ in our analysis \cite{Baak:2014ora}.

In 2HDMs without tree-level FCNCs, flavour constraints are not that significant, except the radiative decay $b \rightarrow s \gamma$, which gives a lower bound of 480 GeV (recently uplifted to 580 GeV \cite{Misiak:2017bgg}) on the charged Higgs mass \cite{Misiak:2015xwa} . This bound holds for type- II and flipped 2HDM only.
\item {\textbf{\em{Alignment limit :}}} \\
The parameter space is also compatible with the "alignment limit" at which  one of the Higgs mass eigenstates becomes identical with SM
Higgs boson and the couplings with gauge bosons and fermions become SM-like. We have tried to set $\beta - \alpha \sim \frac{\pi}{2}$ throughout by imposing proper limits on cos($\beta-\alpha$).
\item {\textbf{\em{Collider constraints :}}} \\
 The mixing angles $\alpha$, $\beta$ as well as different couplings are tightly constrained from the measured signal strengths for different channels, which takes us close to the alignment limit. $2 \sigma $ allowed ranges for all the signal strengths have been used throughout our analysis \cite{CMS-PAS-HIG-15-002}.
\end{enumerate}

\subsubsection{Veltman conditions for 2HDMs with $CP$-conserving  potential}
We split the VCs into bosonic and fermionic parts as 
\bea
VC_{\Phi} &=& VC^B_{\Phi} + VC^F_{\Phi}.
\eea
The functional forms of the bosonic part of the VCs are independent 
of the $\mathbb{Z}_2$ charges of the component fields.
\besub
\bea
VC^B_{h} &=& \Big(\l_1 \text{sin}^2 \a + \l_2 \text{cos}^2 \a + 2\l_3 + \l_4 - 3(\l_6 + \l_7) 
\text{sin}2\a + \frac{3}{4}g_1^2 + \frac{9}{4}g_2^2\Big) \\
VC^B_{H} &=& \Big(\l_1 \text{cos}^2 \a + \l_2 \text{sin}^2 \a + 2\l_3 + \l_4 + 3(\l_6 + \l_7) 
\text{sin}2\a + \frac{3}{4}g_1^2 + \frac{9}{4}g_2^2\Big) \\
VC^B_{A} &=& \Big(\l_1 \text{sin}^2 \b + \l_2 \text{cos}^2 \b + 2\l_3 + \l_4 + 3(\l_6 + \l_7) 
\text{sin}2\b + \frac{3}{4}g_1^2 + \frac{9}{4}g_2^2\Big) \\
VC^B_{H^\pm} &=& \Big(\l_1 \text{sin}^2 \b + \l_2 \text{cos}^2 \b + 2\l_3 + \l_4 + 3(\l_6 + \l_7) 
\text{sin}2\b + \frac{3}{4}g_1^2 + \frac{9}{4}g_2^2\Big) 
\eea
\eesub
The fermionic part of the VC of a scalar $S (= h, H, A, H^\pm)$ is given by
\bea
VC^F_{S} &=& - 12 \Big(\frac{m_t}{v} f^S_t\Big)^2
- 12 \Big(\frac{m_b}{v} f^S_b\Big)^2
 - 4 \Big(\frac{m_{\tau}}{v} f^S_{\tau}\Big)^2
\eea
For type-II 2HDM we have,
\bea
f^h_t = \frac{\cos \alpha}{\sin\beta}, ~ f^h_b = -\frac{\sin \alpha}{\cos\beta}, ~ f^h_\tau = -\frac{\sin \alpha}{\cos\beta} \,.
\eea
$f^S_t, f^S_b, f^S_\tau$ for the scalars other than $h$ and for different types of 2HDM can be found in reference \cite{Branco:2011iw}.
\subsection{$S_3$-symmetric 2HDM}
When an $S_3$ symmetry is imposed on the quartic terms, 
the scalar potential takes a more restrictive form as shown below:
\begin{eqnarray}
V_4(\Phi_1,\Phi_2) &=& \lambda_1 (\Phi_1^\dagger\Phi_1+\Phi_2^\dagger\Phi_2)^2
+\lambda_2 (\Phi_1^\dagger\Phi_2 -\Phi_2^\dagger\Phi_1)^2 \nonumber\\*
&& + \lambda_3
\left\{(\Phi_1^\dagger\Phi_2+\Phi_2^\dagger\Phi_1)^2
+(\Phi_1^\dagger\Phi_1-\Phi_2^\dagger\Phi_2)^2\right\}\,.  
\label{V4}
\end{eqnarray}
The full potential can be written as,
\bea
V(\Phi_1,\Phi_2) = \mu_1^2 (\Phi_1^\dag \Phi_1) + \mu_2^2 (\Phi_2^\dag \Phi_2) - (\mu_{12}^2 (\Phi_1^\dag \Phi_2) + \rm{h.c.}) + V_4(\Phi_1,\Phi_2) \nonumber \\
\label{full_pot}
\eea
Note that in the $\mu_{12} = 0, \mu_{11} = \mu_{22}$ limit, the full scalar potential becomes invariant
under $S_3$. Denoting tan$\beta = \frac{v_2}{v_1}$, the squared scalar masses for this 
scenario are given by
\begin{eqnarray}
\label{m_h} m^2_{h} &=& 2(\l_1 + \l_3) v^2, \\
m^2_{H} &=& \frac{2 \mu^2_{12}}{\text{sin}{2\b}}, \\
m^2_A &=& \frac{2 \mu^2_{12}}{\text{sin}{2\b}} - 2(\l_2 + \l_3) v^2, \\
m^2_{H^\pm} &=& \frac{2 \mu^2_{12}}{\text{sin}{2\b}} - 2\l_3 v^2.
\end{eqnarray}
One must however note that the tadpole equations $\frac{\partial V}{\partial v_1}$ = 0 = 
$\frac{\partial V}{\partial v_2}$ lead to the additional condition tan$\beta = 1$.

The following set of Yukawa interactions are invariant under the gauge and $S_3$
symmetries.
\begin{eqnarray}
 \mathcal{L}_Y^{(u)} &=&
\null - a_u \Big( 
\bar Q_{1L} \tilde\Phi_1 + \bar Q_{2L}\tilde\Phi_2 \Big)
u_{3R} - b_u \Big\{ \Big( \bar Q_{1L}\tilde\Phi_2 + \bar
Q_{2L}\tilde\Phi_1\Big) u_{1R} + 
\Big( \bar Q_{1L}\tilde\Phi_1 -
\bar Q_{2L}\tilde\Phi_2 \Big)u_{2R} \Big\} \nonumber \\*
&& \null \qquad - c_u \bar Q_{3L} \Big( \tilde\Phi_1 u_{1R} +
\tilde\Phi_2u_{2R} \Big) 
 + {\rm h.c.}
\label{uYuk1}
\end{eqnarray}
where $\tilde{\Phi_i} = i \sigma_2 \Phi_i^*$.

Here ${Q}_{1L}, {Q}_{2L}, {Q}_{3L}$ are left-handed quark doublets, $u_{1R}, u_{2R}, u_{3R}$ are right-handed up-type and down-type singlets. $a_u, b_u, c_u$ are Yukawa couplings for up sector. 
Interactions with the down-type quarks can be written similarly using the Yukawa couplings $a_d, b_d$ and $c_d$ and by replacing $u_{iR}$ by $d_{iR}$, $i = 1,2,3$ in Eq.(\ref{uYuk1}).

This gives rise to the following mass matrix for the quarks ($q = u, d$).
\begin{eqnarray}
M_q = \frac v{\surd2}\begin{pmatrix}
b_q \sin\beta  &  b_q \cos\beta & a_q \cos\beta \\
b_q \cos\beta  & - b_q \sin\beta & a_q \sin\beta \\
c_q \cos\beta  & c_q \sin\beta & 0 \\
\end{pmatrix} \,,
\label{Mq}
\end{eqnarray}
Such a texture for $M_q$ makes solving analytically for the eigenvalues difficult.
A possible way out is to opt for the approximation $c_q << a_q,b_q$, and thereby split
the mass matrix in the following fashion 
\bea
M_q = M^0_q + M^{\prime}_q,
\eea
such that $M^{\prime}_q < < M^0_q$. Here,
\begin{eqnarray}
M^0_q = \frac v{\surd2}\begin{pmatrix}
b_q \sin\beta  &  b_q \cos\beta & a_q \cos\beta \\
b_q \cos\beta  & - b_q \sin\beta & a_q \sin\beta \\
0  & 0 & 0 \\
\end{pmatrix} \,,
\label{Mq}
\end{eqnarray}

\begin{eqnarray}
M^{\prime}_q = \frac v{\surd2}\begin{pmatrix}
0  &  0 & 0 \\
0  & 0 & 0 \\
c_q \cos\beta  & c_q \sin\beta & 0 \\
\end{pmatrix} \
\label{Mq}
\end{eqnarray}
The matrix $M^0_q$ is non-hermitian and thus, can be brought to a diagonal form
using a bi-unitary transformation. However, one of its eigenvalues
remains identically zero, thereby making it impossible to reconstruct the quark
mass hierarchy using $M^0_q$ alone. One way to resolve this is to treat $M^{\prime}_q$
as a perturbation to $M^0_q$ and compute the correction to the eigenvalues using first
order (non-degenerate) perturbation theory. It turns out that the zero eigenvalue shifts
to $-\frac{v}{ 2}\frac{a c}{a^2 + b^2}$.

%
%

%
%

The quark masses then can be uniquely reconstructed as $m_{u(d)} = \frac{1}{\sqrt 2} v c_{u(d)}, m_{c(s)} = \frac{1}{\sqrt 2} v a_{u(d)}, m_{t(b)} = \frac{1}{\sqrt 2} v \sqrt{a_{u(d)}^2 + b_{u(d)}^2}$. This uniquely
fixes $a_{u(d)}, b_{u(d)}$ and $c_{u(d)}$. The VCs in this case then 
take the following forms
\bea
VC_{h,H} &=& -10.8038 + \frac{3}{4} g_1^2 + \frac{9}{4} g_2^2 + 28 \lambda_1 - 5 \lambda_2 + 15 \lambda_3 \,, \nonumber \\
VC_A &=& -10.8038 + \frac{3}{4} g_1^2 + \frac{9}{4} g_2^2 + 5.5 \lambda_1 -  \lambda_2 + 2.5 \lambda_3 \,, \nonumber \\
VC_{H^\pm} &=& -21.6076 + \frac{3}{4} g_1^2 + \frac{9}{4} g_2^2 + 11 \lambda_1 - 2 \lambda_2 + 4 \lambda_3\,.
\label{VC_S3_2HDM}
\eea

\subsection{Results}
\label{results_2HDM}
We describe the key features of our results in this subsection.

In the four canonical 2HDMs, it is not possible to put bounds on the
individual masses owing to the presence of the free parameter $m_{12}$ (Eq.(\ref{2HDM_pot})). Since no fermions couple to $\Phi_1$ in type-I, 
the VCs of the non-standard scalars do not experience the necessary 
negative fermionic contribution. The quartic couplings also cannot be appropriately negative since this violates the stability conditions. Therefore, no viable parameter space exists in this case that keeps quadratic divergence of all the scalars at a manageable level.  

\begin{figure} 
\begin{center}
%
\includegraphics[scale=0.38]{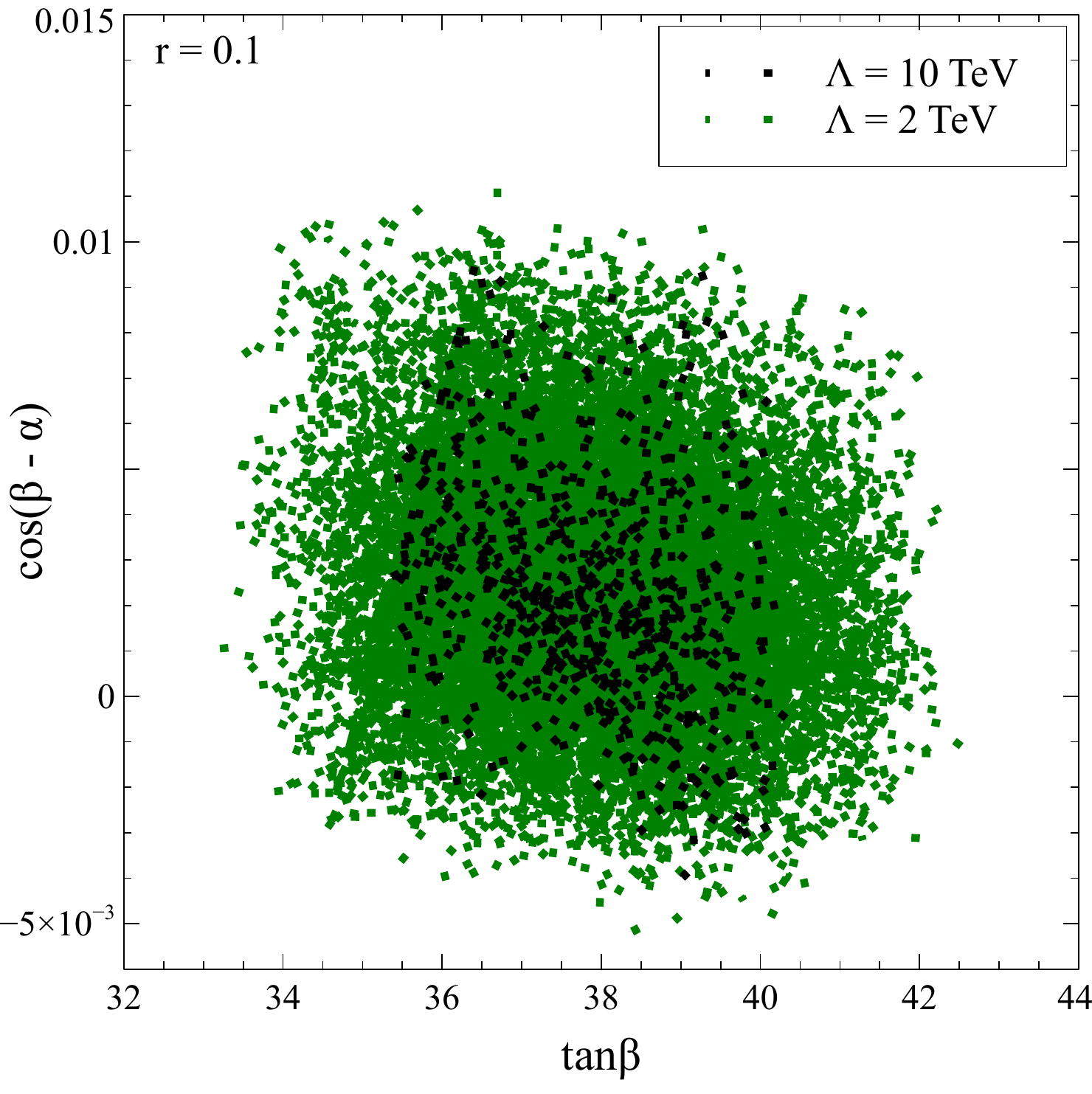}~~~ 
\includegraphics[scale=0.38]{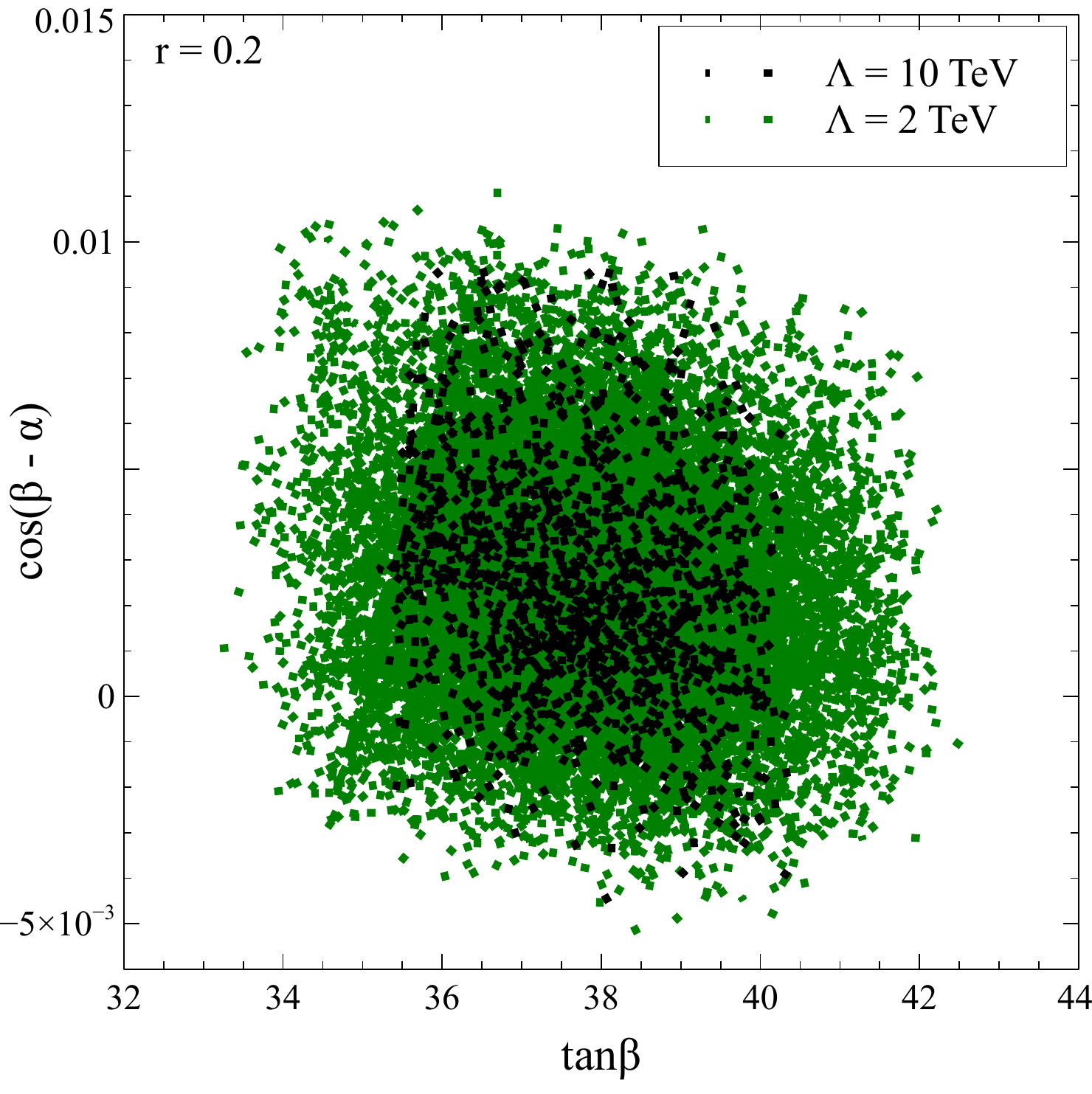}~~~ 
\includegraphics[scale=0.38]{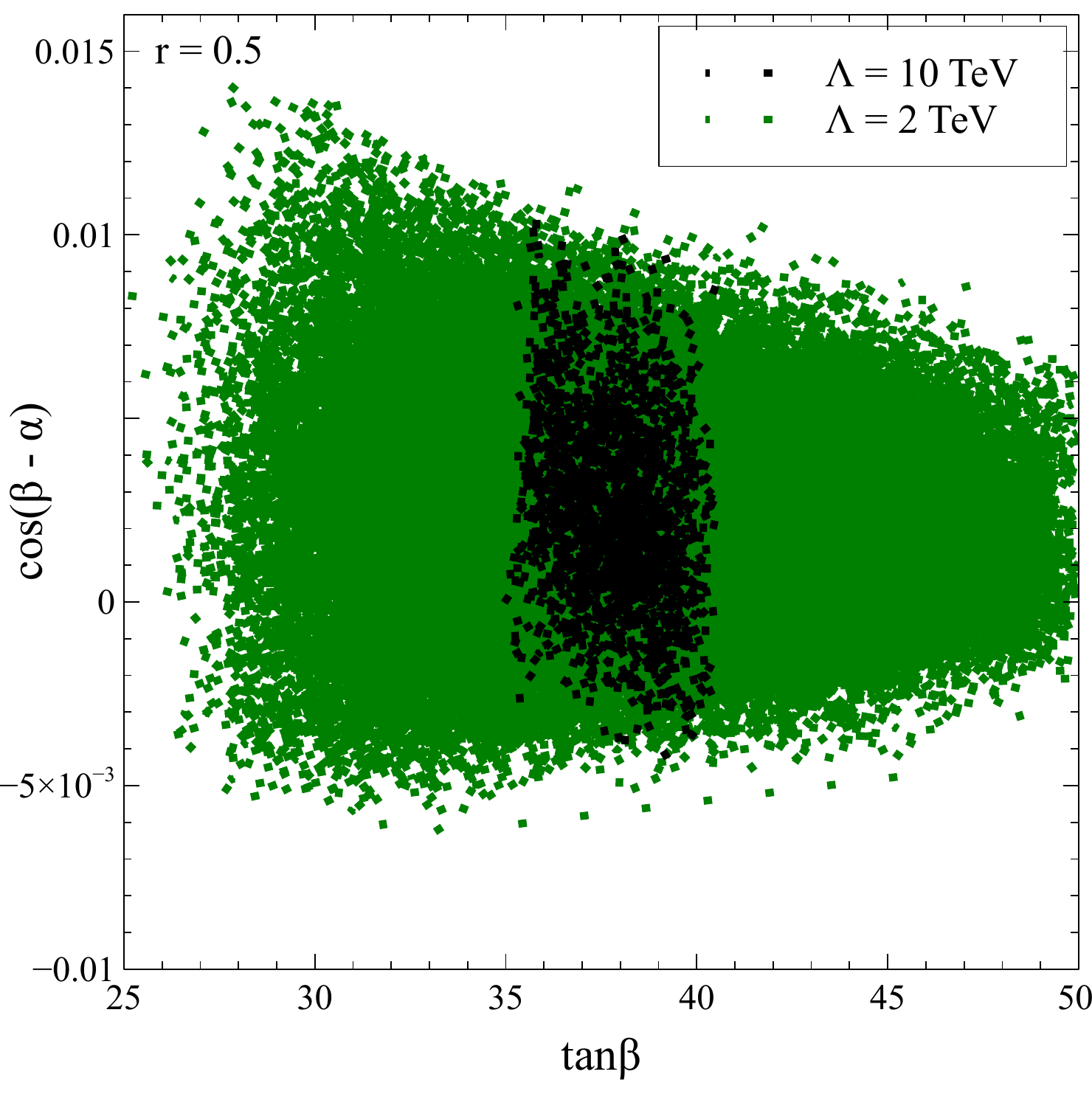}~~~ 
\\
\includegraphics[scale=0.38]{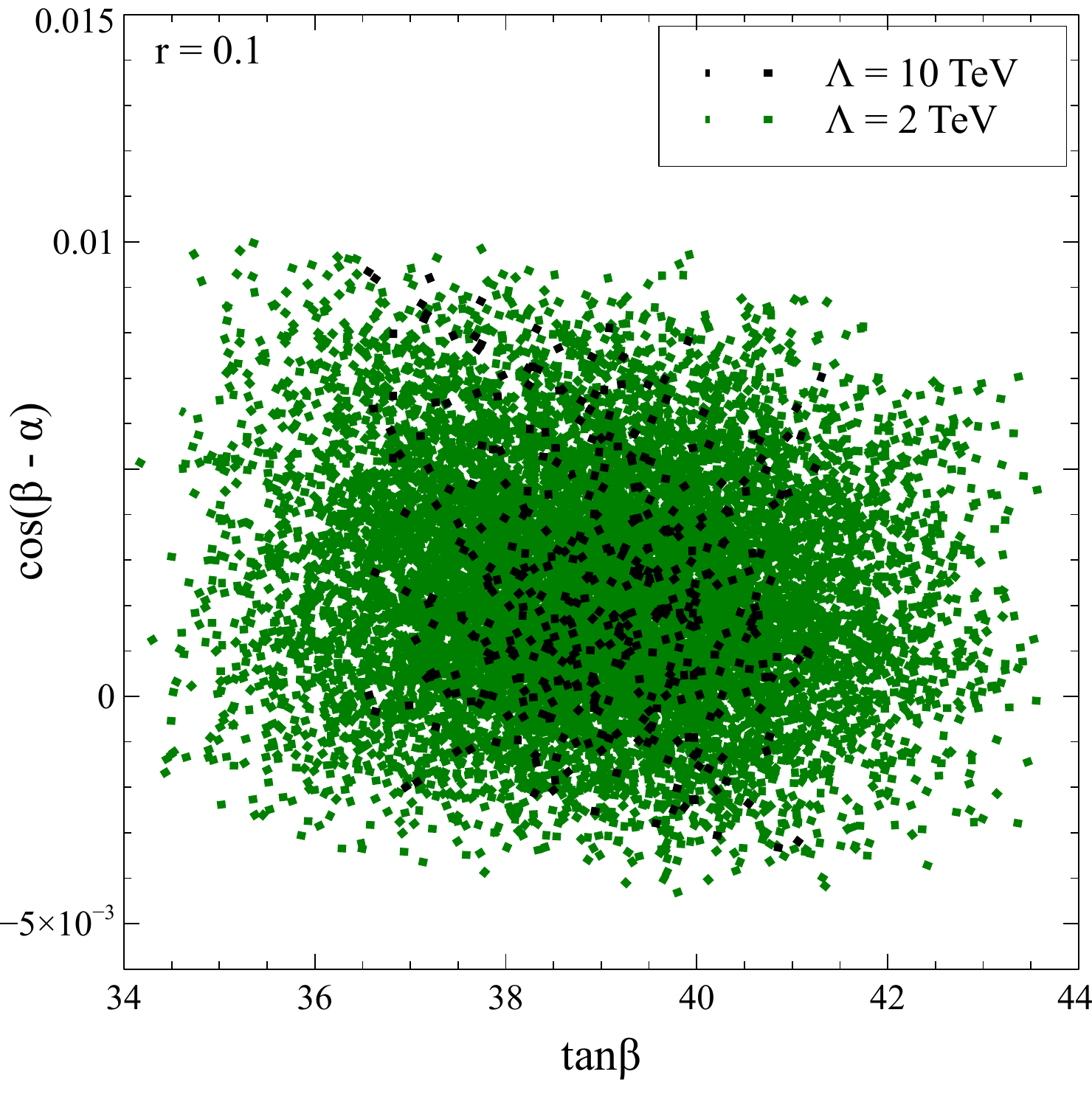}~~~ 
\includegraphics[scale=0.38]{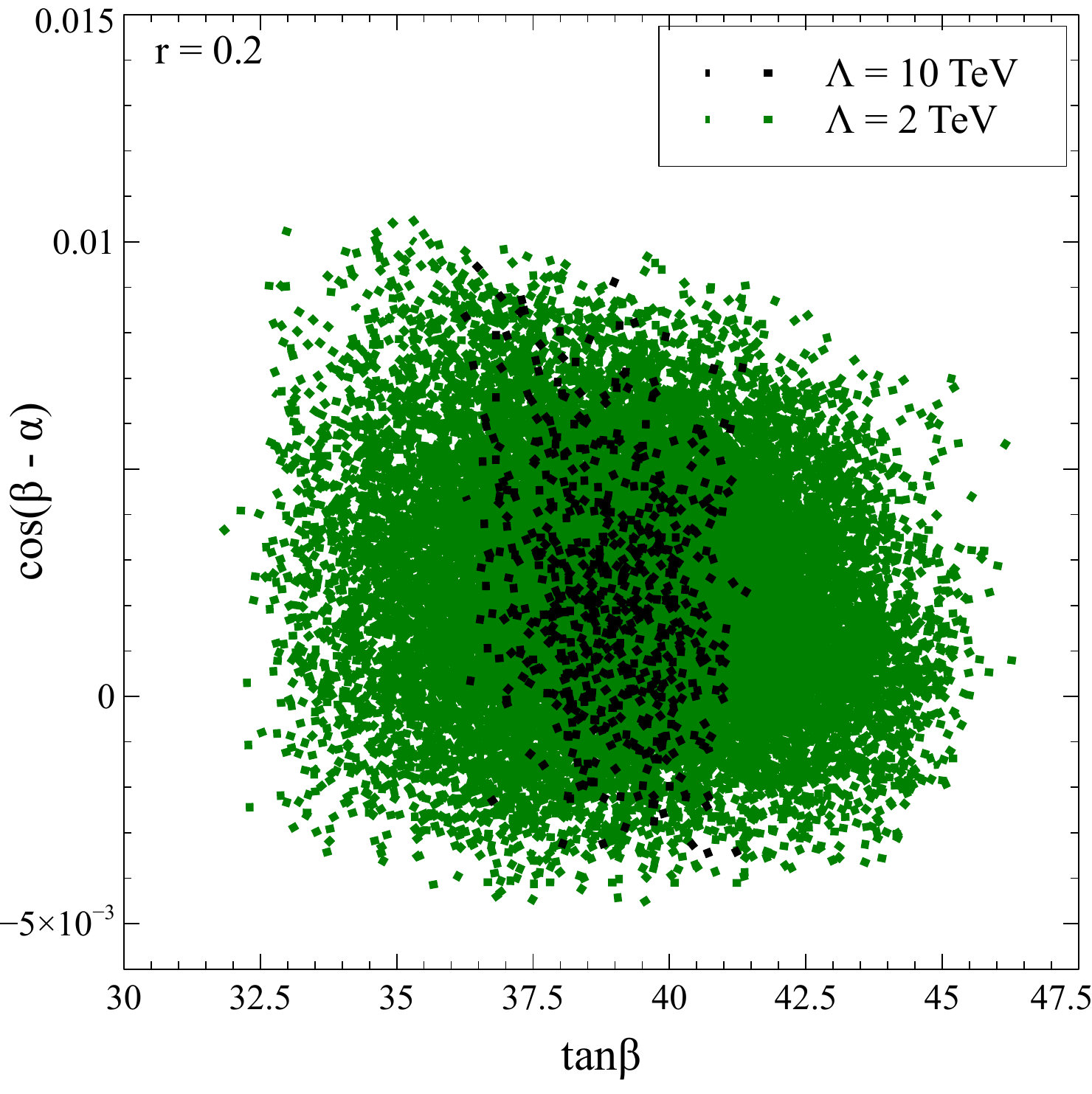}~~~ 
\includegraphics[scale=0.38]{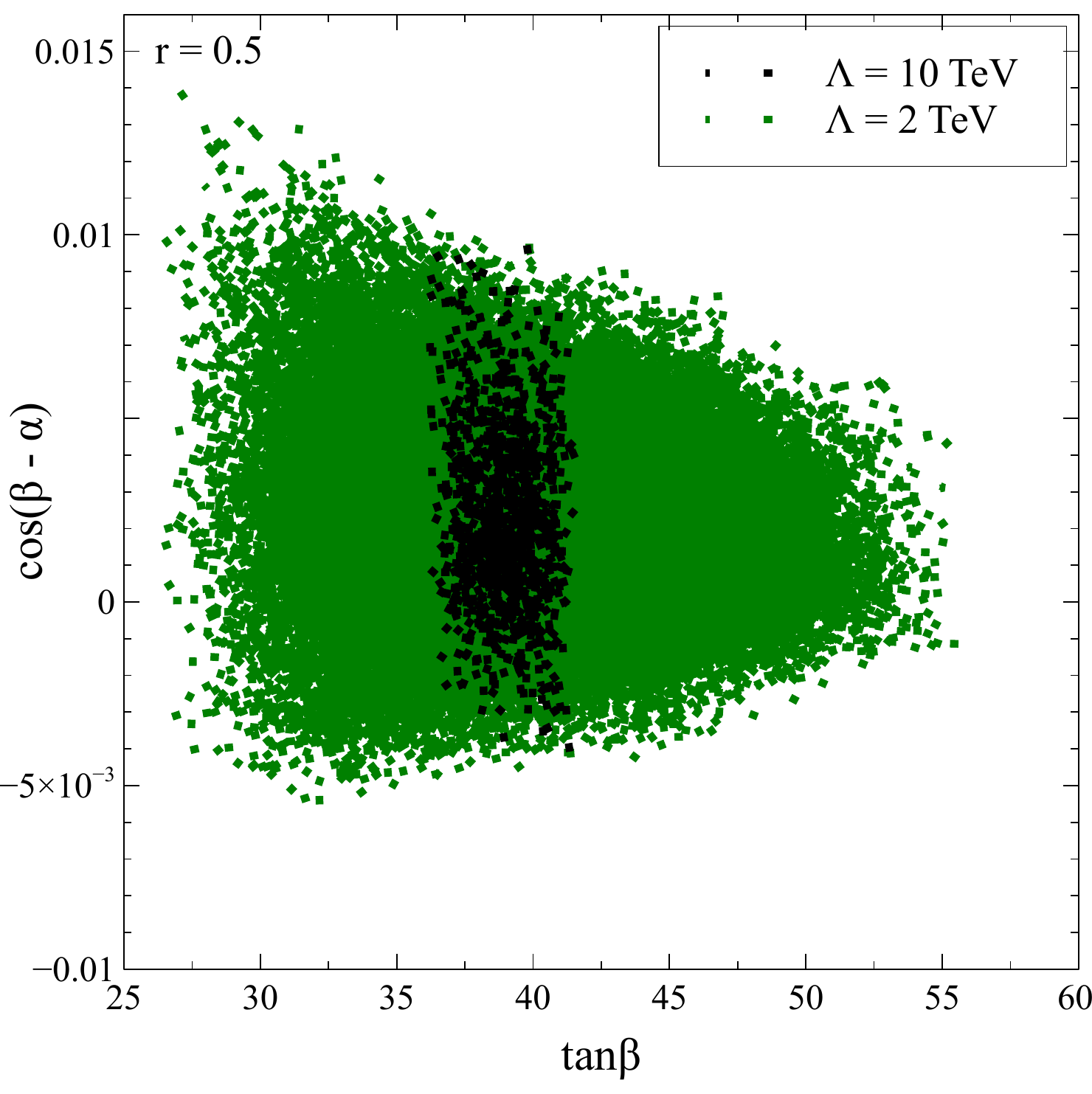}~~~ 
\\
\caption{Allowed parameter spaces in the 
tan$\beta$ vs cos($\b-\a$) planes for $\L = 2 ~\text{TeV},~10$ TeV and $r = 0.1, ~0.2, ~0.5$ for type-II (upper panel) and flipped 2HDMs (lower panel).
The colour coding is explained in the legends. Here, for instance, $r = 0.1$ refers to $|r_{\phi}|$ < 0.1.}
\label{f:masses_2hdm}
\end{center}
\end{figure}

For type-II, the couplings of the $b$-quark with $A,H,H^\pm$ become comparable
in size with the corresponding $t$-Yukawa couplings for high tan$\beta$. This gives a handle to control all the VCs. The scan results can be seen in Fig.\ref{f:masses_2hdm}, where the parameter points clearing the VC constraints are plotted in the tan$\beta$ versus cos$(\b-\a)$ plane for multiple values
of $r$ and $\L$. 

For example, for $\L = 2$ TeV and $|r_{\phi}|$ < 0.1, we obtain 33 < tan$\beta$ < 43 and -0.005 < cos$(\b - \a)$ < 0.01. It is noted that the bound
on cos$(\b - \a)$ is stronger than what is obtained by the imposition of the 
signal-strength constraints alone. The bounds get tighter upon taking $\L = 10$ TeV quite expectedly. The results for the flipped case are similar to type-II, since their Yukawa couplings differ only through the $\tau$. In the flipped case, the allowed range of tan$\beta$ consequently is slightly shifted towards the higher side with respect to the type-II range. One does not
observe manageable quadratic divergences in the non-standard masses in the
Lepton-specific case. 
Relevant discussions on the radiative stability of the 2HDM VCs are in \cite{Chakraborty:2014oma}.

All that is left to discuss in this section is the extent of fine-tuning achievable for the 
IDM as well as the $S_3$-symmetric 2HDM. 
\subsubsection{Results for IDM}
\label{results_IDM}
In case of the IDM, the VC corresponding to the 
$\mathbb{Z}_2$-odd scalars $H,A,H^\pm$ does not carry fermionic terms. Therefore, one has to
rely on possible cancellations amongst the bosonic terms themselves in order to have fine-tuning. 
It is found that the cancellations can at most be partial. This is attributed not only to the 
stability constraints discussed in the previous section, but also to the requirement of a vanishing
Veltman coefficient for the 125 GeV Higgs. 

We have fixed $\l_1 = \frac{m^2_h}{v^2}$ and varied the other quartic couplings in the range $|\l_2|,|\l_4|,|\l_5| < 4\pi$, along with $0 < \mu_{22} < 300$ GeV. In addition, $\l_3$ is chosen so as to make $VC_h = 0$. The lowest $VC_{H}$ thus obtainable in this case is $\simeq 9.1$, thereby leading to $\delta {m_H^2} \simeq (250 ~\rm GeV)^2$ for the cutoff at 1 TeV.
\begin{figure} 
\begin{center}
%
\includegraphics[scale=0.38]{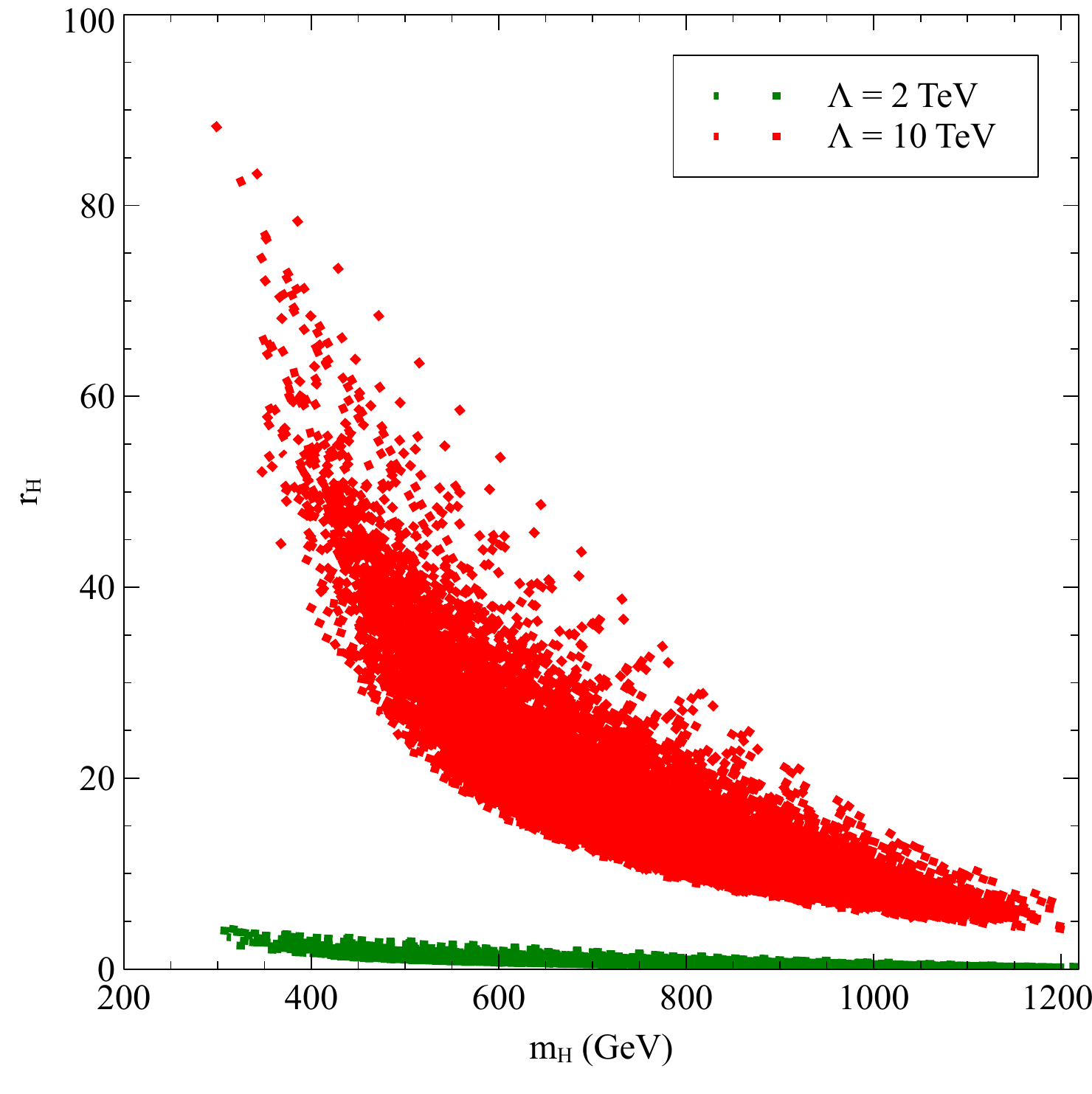}~~~ 
\caption{$r_H$ plotted versus $m_H$ for $\L = $ 2 TeV (green) and $\L = 10$ TeV (red).}
\label{f:rH-mH_idm}
\end{center}
\end{figure}
Therefore, fine-tuning in the IDM fares worse compared to
the type-II 2HDM for instance. This is only expected, keeping in mind the very non-democratic nature of 
Yukawa interactions in this scenario. We define $r_{\Phi} = \frac{\delta {m_{\Phi}^2}}{m^2_{\Phi}}$, where, $\Phi$
refers to $H,A,H^\pm$. As shown in Fig.\ref{f:rH-mH_idm}, it becomes thus possible to restrict $r_{\Phi}$ to small values by opting for larger tree-level masses. Also, in absence of complete cancellation of the VCs in the
inert sector, increasing the cutoff $\L$ only worsens the fine-tuning.

\subsubsection{Results for $S_3$-symmetric 2HDM}
The Yukawa couplings of this model being completely fixed by the quark masses, the fermionic contributions to the Veltman coefficients in Eq.(\ref{VC_S3_2HDM}) are fixed for the particular choice of $\tan \beta$ we have made, {\em i.e. $\tan \beta = 1$}. We have scanned over $\lambda_1$ and  $\mu_{12}^2$ within the following ranges :
\bea
0.0 < \lambda_1 < 1.0 ,~ 0.0~{\rm GeV^2}  < \mu_{12}^2 < 10^6 ~{\rm GeV^2} \,.
\eea
Given the SM Higgs mass $m_h = 125$ GeV, $\lambda_3$ has been fixed using Eq.(\ref{m_h}). $\lambda_2$ has been chosen in such a way that the Veltman conditions for the SM Higgs boson and the Heavy Higgs boson (since the Veltman conditions for $h$ and $H$ are same) are satisfied exactly.

With this set of parameters satisfying the Veltman condition for SM Higgs boson and Heavy Higgs, it is not possible to satisfy the Veltman conditions for pseudoscalar Higgs $A$ and charged Higgs $H^\pm$ simultaneously. For these two mass eigenstates, we adjust the ratio $r_{\Phi}$ for a fixed value of $\Lambda = 1$ TeV, mentioned in \ref{results_IDM}, so that $\frac{\delta m_\Phi^2}{m_\Phi^2} < 1$ , where $\Phi = A, H^\pm$ in this context. The minimum value of $r_\Phi$ required to satisfy all other constraints is 0.1 at $\Lambda = 1$ TeV. Increasing the value of $r_\Phi$ will enhance the parameter space. With increasing $\Lambda$, minimum value of $r_\Phi$ which
is required to comply with the existing constraints, increases. 

To conclude this section, we remark that the type-II and flipped 2HDM
to be most favourable from the point of view of Higgs mass fine-tuning, 
amongst the cases considered here. As highlighted before, the key to these findings lies in the Yukawa interactions . Hence, an interesting task would be to probe fine-tuning in a 2HDM with a more generic Yukawa structure, modulo constraints from flavour physics.

\section{Three Higgs doublet models}
\label{3HDM}
We now move on to the next simplest example, {\em i.e.} the 3HDMs \cite{Keus:2013hya} in order to address the fine-tuning problem. The main motivation of the 3HDMs is that the three fermion families can be connected with the three scalar doublets through appropriate symmetries so that the observed pattern of the fermion masses and
mixings is reproduced. In addition, the scalar sector of a 3HDM is richer than compared to a 2HDM. The additional quartic couplings present in the former can lead to interesting constraints \emph{vis-a-vis} fine-tuning.
With all possible finite symmetries being clearly identified in 3HDMs \cite{Ivanov:2012fp}, the flavour problem can also be addressed by this extension. The classification of 3HDMs can be made in terms of all possible continuous and discrete abelian symmetries as well as all possible discrete non-abelian symmetries \cite{Keus:2013hya}.

The present study contains 3HDMs with the discrete symmetries $A_4$, $S_4$ and $S_3$ described in the following subsections. With these types of 3HDMs, we shall explore the degree of satisfaction of the Veltman conditions corresponding to the different scalars of the 3HDM.
\subsection{$A_4$-symmetric 3HDM}
\subsubsection{Potential of $A_4$-symmetric 3HDM}
$A_4$-symmetric 3HDM scalar potential can be written as \cite{Degee:2012sk,Ivanov:2014doa},
\bea
V (\Phi) &=& -\frac{M_0}{\sqrt{3}} \left(\Phi_1^\dag \Phi_1 + \Phi_2^\dag \Phi_2+ \Phi_3^\dag \Phi_3\right) + \frac{\lambda_0}{3} \left(\Phi_1^\dag \Phi_1 + \Phi_2^\dag \Phi_2+ 
\Phi_3^\dag \Phi_3\right)^2  \nonumber \\
&& + \frac{\lambda_3}{3} [\left(\Phi_1^\dag \Phi_1\right)^2 + \left(\Phi_2^\dag \Phi_2\right)^2 + \left(\Phi_3^\dag \Phi_3\right)^2 - \left(\Phi_1^\dag \Phi_1\right) \left(\Phi_2^\dag \Phi_2\right) \nonumber \\
&& - \left(\Phi_2^\dag \Phi_2\right) \left(\Phi_3^\dag \Phi_3\right) - \left(\Phi_3^\dag \Phi_3\right) \left(\Phi_1^\dag \Phi_1\right)]  \nonumber \\
&& + \lambda_1 \left[\left(\rm{Re}\left(\Phi_1^\dag \Phi_2\right)\right)^2 + \left(\rm{Re}\left(\Phi_2^\dag \Phi_3\right)\right)^2 + \left(\rm{Re}\left(\Phi_3^\dag \Phi_1\right)\right)^2 \right] \nonumber \\
&& + \lambda_2 \left[\left(\rm{Im}\left(\Phi_1^\dag \Phi_2\right)\right)^2 + \left(\rm{Im}\left(\Phi_2^\dag \Phi_3\right)\right)^2 + \left(\rm{Im}\left(\Phi_3^\dag \Phi_1\right)\right)^2 \right] \nonumber \\
&& + \lambda_4 [\left(\rm{Re}\left(\Phi_1^\dag \Phi_2\right)\right)\left(\rm{Im}\left(\Phi_1^\dag \Phi_2\right)\right) + \left(\rm{Re}\left(\Phi_2^\dag \Phi_3\right)\right)\left(\rm{Im}\left(\Phi_2^\dag \Phi_3\right)\right) \nonumber \\
&& + \left(\rm{Re}\left(\Phi_3^\dag \Phi_1\right)\right)\left(\rm{Im}\left(\Phi_3^\dag \Phi_1\right)\right) ] \,.
\label{potential_A4}
\eea
where each doublet is defined as,
\bea
\Phi_i = \begin{pmatrix}
w_i^+ \\
\frac{(h_i + v_i + i z_i)}{\sqrt{2}}
\end{pmatrix} , 
i = 1,2,3
\eea
Here the parameters $M_0$, $\Lambda_i$ are generic real parameters.
The potential mentioned above possesses an $A_4$-symmetry, {\em i.e.} invariant under the cyclic permutation of three doublets $\Phi_1, \Phi_2, \Phi_3$. In addition, the potential is also symmetric under independent sign flips of the individual doublets and specific type of generalised $CP$  transformation. Thus as a whole the potential is symmetric under a larger group, {\em i.e.} $A_4 \times Z_2$.

Possible VEV structures of $A_4$-symmetric potential which aid global minima are \cite{Felipe:2013ie}, 
\begin{itemize}
 \item $v(1,0,0)$
 \item $v(1,1,1)$
 \item $v(\pm 1, \eta , \eta^*)$ with $\eta = e^{\frac{i\pi}{3}}$
 \item $v(1, e^{i \alpha}, 0)$ for an arbitrary $\alpha$.
\end{itemize}
Amongst the above, the first and fourth cases do not lead to non-zero fermion
masses for all three generations, and thus, are not phenomenologically viable. In addition, we also do not invoke spontaneous $CP-$violation in our analysis and thus, we do not take this case for further consideration.
This leaves us the second VEV configuration {\em i.e.} $v(1,1,1)$ to illustrate our results. We write down the corresponding Yukawa Lagrangian
in the following subsection.

The basis transformation laws between the flavour eigenstates (the left-hand side of the equations below) and mass eigenstates (the right-hand side of the equations below) of the scalars are given below,
\bea
\begin{pmatrix}
h_1 (z_1) \\
h_2 (z_2) \\
h_3 (z_3)
\end{pmatrix} = \begin{pmatrix}
                \frac{1}{\sqrt{3}} & \frac{1}{\sqrt{3}} & \frac{1}{\sqrt{3}} \\
                \frac{1}{\sqrt{2}} & 0 & -\frac{1}{\sqrt{2}} \\
                \frac{1}{\sqrt{6}} & -\frac{2}{\sqrt{6}} & \frac{1}{\sqrt{6}}
                 \end{pmatrix}
                 \begin{pmatrix}
                 h (A_3) \\
                 H_1 (A_1) \\
                 H_2 (A_2)
                 \end{pmatrix}
\eea

The physical scalar spectrum of $A_4$-symmetric 3HDM consists of three neutral scalars $h, H_1, H_2$ (lighter mass eigenstate $h$ identified as Standard Model Higgs boson with mass 125 GeV), two pseudoscalars $A_1, A_2$ and two charged scalars $H_1^+, H_2^+$. Mass squared of these physical states can be expressed in terms of the couplings $\lambda_i$'s of the scalar potential and VEV $v = 246$ GeV as,
\bea
m_h^2 &=& \frac {2} {3} (\lambda_0 + \lambda_1)v^2 \,, \nonumber \\
m_{H_1/H_2}^2 &=& \frac 1 3 (-\lambda_1 + \lambda_3)v^2 \,,  \nonumber \\
m_{A_1/A_2}^2 &=& \frac 1 3 (-\lambda_1 + \lambda_2)v^2 \,,  \nonumber \\
m_{H_1^+}^2 &=& \frac {1} {12} (- 6 \lambda_1 - \sqrt{3} \lambda_4)v^2 \,,\nonumber \\
m_{H_2^+}^2 &=& \frac {1} {12} (- 6 \lambda_1 + \sqrt{3} \lambda_4)v^2 \,.
\eea
The scalars in the mass basis are related to the ones in the gauge
basis as
\bea
\begin{pmatrix}
w_1^+ \\
w_2^+ \\
w_3^+ 
\end{pmatrix} = \begin{pmatrix}
                1 & 1 & 1 \\
                (-\frac{\sqrt{3}}{6} - \frac i 2) &  (-\frac{\sqrt{3}}{6} +   \frac i 2) & \frac{1}{\sqrt{3}} \\
                (-\frac{\sqrt{3}}{6} + \frac i 2) &  (-\frac{\sqrt{3}}{6} -   \frac i 2) & \frac{1}{\sqrt{3}}
                 \end{pmatrix}
                 \begin{pmatrix}
                 H_3^+ \\
                 H_1^+ \\
                 H_2^+
                 \end{pmatrix}
\eea
$A_3$ and $H_3^+$ above are identified as neutral and charged Goldstone bosons.

\subsubsection{Yukawa Lagrangian for the VEV configuration $v(1,1,1)$ }
Since three doublets are taken to be in a triplet representation {\bf 3}, to make the Yukawa Lagrangian in the invariant singlet representation {\bf 1}, left-handed fermions and the right-handed fermions should belong to triplet representation {\bf 3} and singlet representation {\bf 1}. This is one of the several choices as mentioned in \cite{Felipe:2013ie} and we shall follow this configuration to construct Yukawa Lagrangian.

Product of two triplets $ x = (x_1, x_2, x_3)$ and $ y = (y_1, y_2, y_3)$ gives singlet representation {\bf 1}, {\bf 1'}, {\bf 1"} as,
\bea
(x\otimes y )_{\bf 1} &=& x_1 y_1 + x_2 y_2 + x_3 y_3 \nonumber \\
(x\otimes y )_{\bf 1'} &=& x_1 y_1 + \omega^2 x_2 y_2 + \omega x_3 y_3 \nonumber \\
(x\otimes y )_{\bf 1"} &=& x_1 y_1 + \omega x_2 y_2 + \omega^2 x_3 y_3 
\label{representations}
\eea

 Thus combining left-handed fermions in the triplet representation {\bf 3}, three Higgs doublets in the triplet representation {\bf 3}, right-handed fermions in the singlet representation {\bf 1} and adopting the VEV structure $v(1,1,1)$, one can write down the $A_4$-symmetric Yukawa Lagrangian providing non-zero and non-vanishing quark masses as follows \cite{Felipe:2013ie},
\bea
\mathcal{L}_{Yukawa} &=& \alpha_1 (\bar{Q}_{1L} \tilde{\Phi_1} + \bar{Q}_{2L} \tilde{\Phi_2} + \bar{Q}_{3L} \tilde{\Phi_3}) u_{1R} +  \alpha_2 (\bar{Q}_{1L} \tilde{\Phi_1} + \omega ~\bar{Q}_{2L} \tilde{\Phi_2} + \omega^2 ~\bar{Q}_{3L}  \tilde{\Phi_3}) u_{2R} \nonumber \\
&& + \alpha_3 (\bar{Q}_{1L} \tilde{\Phi_1} + \omega^2 ~ \bar{Q}_{2L} \tilde{\Phi_2} + \omega ~\bar{Q}_{3L} \tilde{\Phi_3}) u_{3R} + 
\beta_1 (\bar{Q}_{1L} \Phi_1 + \bar{Q}_{2L} \Phi_2 + \bar{Q}_{3L} \Phi_3) d_{1R} \nonumber \\
&& +  \beta_2 (\bar{Q}_{1L} \Phi_1 + \omega ~\bar{Q}_{2L} \Phi_2 + \omega^2 ~\bar{Q}_{3L}  \Phi_3) d_{2R} + \beta_3 (\bar{Q}_{1L} \Phi_1 + \omega^2 ~ \bar{Q}_{2L} \Phi_2 + w ~\bar{Q}_{3L} \Phi_3) d_{3R} \nonumber\\
&& + {\rm h.c.} \,,
\label{Yukawa_A4}
\eea
with $\omega = e^{\frac{2i\pi}{3}}$, $\tilde{\Phi_i} = i \sigma_2 \Phi_i^*$.

Here ${Q}_{1L}, {Q}_{2L}, {Q}_{3L}$ are left-handed quark doublets, $u_{1R}, u_{2R}, u_{3R}, d_{1R}, d_{2R}, d_{3R}$ are right-handed up-type and down-type singlets. [$\alpha_1, \alpha_2, \alpha_3$], [$\beta_1, \beta_2, \beta_3$] are Yukawa couplings for up-type and down-type sectors respectively.

Left-handed and right-handed up-type and down-type fermions are transformed from their flavour eigenstates to mass eigenstates following the transformation rules given below.

\bea
\begin{pmatrix}
u_{1L/R} (d_{1L/R})\\
u_{2L/R} (d_{2L/R}) \\
u_{3L/R} (d_{3L/R})
\end{pmatrix} = U_{L/R}(V_{L/R})
\begin{pmatrix}
u_{L/R} (d_{L/R})\\
c_{L/R} (s_{L/R}) \\
t_{L/R} (b_{L/R})
\end{pmatrix}
\eea
Electroweak symmetry breaking gives rise to the mass matrix in the up-type quark sector,
\bea
M_u = \frac{1}{\sqrt{6}}\begin{pmatrix}
                 v \alpha_1 & v \alpha_1 & v \alpha_1 \\
                 v \alpha_2 & (-\frac 1 2 + \frac{i \sqrt{3}}{2})v \alpha_2 & (-\frac 1 2 - \frac{i \sqrt{3}}{2})v \alpha_2 \\
                 v \alpha_3 & (-\frac 1 2 - \frac{i \sqrt{3}}{2})v \alpha_3 & (-\frac 1 2 + \frac{i \sqrt{3}}{2})v \alpha_3 
                  \end{pmatrix}
\label{mass_matrix_up}
\eea
Similarly mass matrix $M_d$ can be written for the down-type sector by replacing $\alpha_i$'s with $\beta_i$'s in Eq.(\ref{mass_matrix_up}). $M_u (M_d)$ can be diagonalised by bi-unitary transformations to give quark masses as,
\bea
M_u^{diag} = U_L^{-1} M_u U_R  \,,
~ M_d^{diag} = V_L^{-1} M_d V_R \,.
\label{M_diag}
\eea
where,
\bea
U_L(V_L) = \begin{pmatrix}
                 1 & 0 & 0 \\
                 0 & 1 & 0 \\
                 0 & 0 & 1 
                  \end{pmatrix} \,, 
~ U_R(V_R) =  \begin{pmatrix}
                 \frac{1}{\sqrt{3}} &  \frac{1}{2\sqrt{3}}(-1 - i \sqrt{3})
                  & \frac{1}{2\sqrt{3}}(-1 + i \sqrt{3}) \\
                   \frac{1}{\sqrt{3}} &  \frac{1}{2\sqrt{3}}(-1 + i \sqrt{3})
                  & \frac{1}{2\sqrt{3}}(-1 - i \sqrt{3}) \\
                   \frac{1}{\sqrt{3}} & \frac{1}{\sqrt{3}} & \frac{1}{\sqrt{3}}
                  
                 \end{pmatrix}                 
\eea
This will yield quark masses as a function of $\alpha_i$'s as given below, apart from a phase factor which can be absorbed.
\besub
\bea
m_u = \frac{v \alpha_1}{\sqrt{2}} \,, 
m_c = \frac{v \alpha_2}{\sqrt{2}} \,, 
m_t = \frac{v \alpha_3}{\sqrt{2}} \,. \\
m_d = \frac{v \beta_1}{\sqrt{2}} \,, 
m_s = \frac{v \beta_2}{\sqrt{2}} \,, 
m_b = \frac{v \beta_3}{\sqrt{2}} \,.
\eea
\label{mass_quarks}
\eesub
Thus $\alpha_1, \alpha_2, \alpha_3, \beta_1, \beta_2, \beta_3$ are not free parameters, but get fixed by the quark masses as seen in Eq.(\ref{mass_quarks}).

\subsubsection{Veltman conditions}
In this section we introduce the Veltman coefficients (mentioned in the Introduction) for the mass eigenstates of this model. As shown in the Section \ref{2HDM}, Veltman coefficient of each physical state consists of bosonic (all with positive sign) and fermionic (all with extra negative sign due to fermionic trace) contributions as given below,
\bea
VC_{h} &=&\frac{14 \text{$\lambda_0$}}{3}+\text{$\lambda_1$}+\text{$\lambda _2$}+\frac{2\text{$\lambda_3$}}{3} + \frac{3}{4} g^2_1 + \frac{9}{4} g^2_2 -2 \text{$\alpha_1$}^2-2 \text{$\alpha_2$}^2-2 \text{$\alpha_3$}^2-2 \text{$\beta_1$}^2-2 \text{$\beta_2$}^2 \nonumber \\
&& -2 \text{$\beta _3$}^2  \nonumber \\
VC_{H_1} &=&\frac{14 \text{$\lambda_0$}}{3}+\text{$\lambda_1$}+\text{$\lambda _2$}+\frac{2\text{$\lambda_3$}}{3} + \frac{3}{4} g^2_1 + \frac{9}{4} g^2_2  \nonumber \\
VC_{H_2} &=&\frac{14 \text{$\lambda_0$}}{3}+\text{$\lambda_1$}+\text{$\lambda _2$}+\frac{2\text{$\lambda_3$}}{3} + \frac{3}{4} g^2_1 + \frac{9}{4} g^2_2 - \text{$\alpha_1$}^2- \text{$\alpha_2$}^2- \text{$\alpha_3$}^2- \text{$\beta_1$}^2- \text{$\beta_2$}^2- \text{$\beta _3$}^2 \nonumber \\
VC_{A_1} &=&\frac{14 \text{$\lambda_0$}}{3}+\text{$\lambda_1$}+\text{$\lambda _2$}+\frac{2\text{$\lambda_3$}}{3} + \frac{3}{4} g^2_1 + \frac{9}{4} g^2_2 \nonumber 
\eea
\bea
VC_{A_2} &=&\frac{14 \text{$\lambda_0$}}{3}+\text{$\lambda_1$}+\text{$\lambda _2$}+\frac{2\text{$\lambda_3$}}{3} + \frac{3}{4} g^2_1 + \frac{9}{4} g^2_2 - \text{$\alpha_1$}^2- \text{$\alpha_2$}^2- \text{$\alpha_3$}^2- \text{$\beta_1$}^2- \text{$\beta_2$}^2- \text{$\beta _3$}^2 \nonumber \\
VC_{{H_1}^\pm} &=& \frac{14 \text{$\lambda_0$}}{3}+\text{$\lambda_1$}+\text{$\lambda _2$}+\frac{2\text{$\lambda_3$}}{3} + \frac{3}{4} g^2_1 + \frac{9}{4} g^2_2 -4 \text{$\alpha_1$}^2-4 \text{$\alpha_2$}^2-4 \text{$\alpha_3$}^2-4 \text{$\beta_1$}^2-4 \text{$\beta_2$}^2 \nonumber \\
&& -4 \text{$\beta _3$}^2 \nonumber \\
VC_{{H_2}^\pm} &=& \frac{14 \text{$\lambda_0$}}{3}+\text{$\lambda_1$}+\text{$\lambda _2$}+\frac{2\text{$\lambda_3$}}{3} + \frac{3}{4} g^2_1 + \frac{9}{4} g^2_2 -4 \text{$\alpha_1$}^2-4 \text{$\alpha_2$}^2-4 \text{$\alpha_3$}^2-4 \text{$\beta_1$}^2-4 \text{$\beta_2$}^2 \nonumber \\
&& -4 \text{$\beta _3$}^2 
\label{VC_A4}
\eea
Quadratically divergent self-energy corrections can be written as,
\bea
\delta m_\Phi^2 = \frac{\Lambda^2}{16 \pi^2} VC_\Phi \,, 
\eea
for $\Phi$ denoting $h, H_1, H_2, A_1, A_2, {H_1}^\pm, {H_2}^\pm$.

The $A_4$-symmetry of the Lagrangian forces all the bosonic contributions to the Veltman coefficients to be identical as can be seen from Eq.(\ref{VC_A4}). After introducing the fermionic contributions, they are slightly different, but still some of them look alike : [$VC_{H_1}, VC_{A_1}$], [$VC_{H_2}, VC_{A_2}$], [$VC_{{H_1}^\pm}, VC_{{H_2}^\pm}$]. $VC_{H_1}, VC_{A_1}$ do not have any fermionic contributions since they get cancelled.

\subsubsection{Constraints and analysis}
\label{constraints}
Our next task is to find out the optimally allowed parameter space compatible with some theoretical constraints given below.
\begin{enumerate}
\item {\textbf{\em Perturbativity :}} \\
We have varied $\lambda_i$'s within a narrow range as mentioned below, so that the perturbativity condition ( $|\lambda_i| \leq 4 \pi , i = 0,1,2,...4.$) is satisfied automatically. The perturbative upper bound for the Yukawa couplings are $|y_i| \leq \sqrt{4 \pi}$ \cite{Chakrabarty:2014aya}, which is more conservative but are clearly satisfied in our case, where Yukawa couplings are completely fixed by the masses of the quarks (Eq.(\ref{mass_quarks})).
\bea
0.0 \leq \lambda_0 \leq 1.0 \,, \nonumber \\
-1.0 \leq \lambda_1, \lambda_2, \lambda_3, \lambda_4  \leq 1.0 \,.
\eea
\item {\textbf{\em Stability conditions :}} \\
The requirement that the potential must remain positive (bounded from below) along various directions in the field space in the limit $\phi_i \rightarrow \infty$, gives rise to the following stability conditions for $A_4$-symmetric 3HDMs.
\bea
\lambda_0 + \lambda_3 \geq 0 \,, \nonumber \\
\lambda_0 + \lambda_3 + 3 \lambda_1 \geq 0 \,, \nonumber \\
\lambda_0 + \lambda_3 - 3 \lambda_2 \geq 0 \,, \nonumber \\
\lambda_0 + \lambda_3 \geq 0 \,.
\label{stability_A4}
\eea
\item $h$ should be identified with the Standard Model Higgs boson, we have varied Higgs mass $m_h$ within ($125.0 \pm 2.0$) GeV.
\end{enumerate}

As mentioned earlier, our main focus would be to adjust the bosonic and fermionic contributions in Veltman coefficient of SM Higgs ($VC_h$) in such a way that the Veltman condition is enforced either strictly by making $VC_h = 0$ or make $VC_h$ as tiny as possible so as to have negligible quadratic divergence. Besides we shall also try to adjust the quadratic divergences originating from the other physical scalars to remain within manageable levels. For this purpose we impose the theoretical constraints discussed in the Sec.\ref{constraints} and obtain an optimal parameter space spanned by the model parameters. Following are our main observations :
\begin{itemize}
\item Since the Yukawa couplings $\alpha_i$'s and $\beta_i$'s are fixed by the up-type and down-type quark masses, except $\alpha_3$ and $\beta_3$ rest of them are tiny and their contributions become more negligible when they appear as squared term in the Veltman coefficients. Thus it is impossible for the fermionic contributions to cancel the bosonic contributions exactly, and the dominance of bosonic contribution over the fermionic counter part results in a positive Veltman coefficient $VC_h$.
\item Since it is not possible to enforce Veltman condition strictly for $h$, we try to fine-tune the ratio $r_h = \frac{\delta m_h^2}{m_h^2}$ defined in Sec. \ref{results_IDM}, and claim that the Veltman coefficient $VC_h$ will be adjusted accordingly, so that $\delta m_h^2 < m_h^2 $ condition is satisfied. For $\Lambda = 1$ TeV, we have set the lowest value of $r_h \approx 0.68$ to satisfy the condition for $h$ only.
\item $VC_{H_1}$, $VC_{A_1}$ contains only bosonic part, $g_1, g_2$ being always positive, to achieve exactly zero Veltman coefficient large negative values of scalar couplings are needed, which is forbidden by the stability conditions given in Eq.(\ref{stability_A4}) since it makes the potential unbounded from below.
\item We have checked that it is also difficult to satisfy the Veltman condition exactly for $H_2 (A_2)$ and $H_1^\pm (H_2^\pm)$ due to the same reason mentioned in case of $VC_h$. If we adopt the same prescription as described in point 2, define the ratio $r_\Phi$ for all the physical scalars other than $h$ and want to abide by the relation $\delta m_\Phi^2 < m_\Phi^2 $ for $\Phi = H_1, H_2, A_1, A_2, {H_1}^\pm, {H_2}^\pm$, it can be seen that it is impossible to fine-tune the ratio for these scalars simultaneously. Thus for $h$ only, the quadratically divergent correction term can be adjusted to be tiny by tuning the ratio $r_\Phi$ accordingly.
\end{itemize}
\subsection{$S_4$-symmetric 3HDM}
We would like to make a comment regarding the $S_4$-symmetric 3HDM which we exclude from our analysis.

The scalar potential of $S_4$-symmetric 3HDM can be written by putting $\lambda_4 = 0$ in Eq.(\ref{potential_A4}) as \cite{Degee:2012sk,Ivanov:2014doa},
\bea
V (\phi) &=& -\frac{M_0}{\sqrt{3}} \left(\Phi_1^\dag \Phi_1 + \Phi_2^\dag \Phi_2+ \Phi_3^\dag \Phi_3\right) + \frac{\lambda_0}{3} \left(\Phi_1^\dag \Phi_1 + \Phi_2^\dag \Phi_2+ 
\Phi_3^\dag \Phi_3\right)^2  \nonumber \\
&& + \frac{\lambda_3}{3} [\left(\Phi_1^\dag \Phi_1\right)^2 + \left(\Phi_2^\dag \Phi_2\right)^2 + \left(\Phi_3^\dag \Phi_3\right)^2 \nonumber \\
&& - \left(\Phi_1^\dag \Phi_1\right) \left(\Phi_2^\dag \Phi_2\right)
- \left(\Phi_2^\dag \Phi_2\right) \left(\Phi_3^\dag \Phi_3\right) - \left(\Phi_3^\dag \Phi_3\right) \left(\Phi_1^\dag \Phi_1\right)]  \nonumber \\
&& + \lambda_1 \left[\left(\rm{Re}\left(\Phi_1^\dag \Phi_2\right)\right)^2 + \left(\rm{Re}\left(\Phi_2^\dag \Phi_3\right)\right)^2 + \left(\rm{Re}\left(\Phi_3^\dag \Phi_1\right)\right)^2 \right] \nonumber \\
&& + \lambda_2 \left[\left(\rm{Im}\left(\Phi_1^\dag \Phi_2\right)\right)^2 + \left(\rm{Im}\left(\Phi_2^\dag \Phi_3\right)\right)^2 + \left(\rm{Im}\left(\Phi_3^\dag \Phi_1\right)\right)^2 \right] \,.
\label{potential_S4}
\eea
In reference \cite{Felipe:2013ie} full mass spectrum of quarks for different VEV configurations like $v(1,0,0), v(1,1,1), v(1,\eta,\eta^*), v(1,i,0)$ are given in detail. It can be seen that these VEV configurations either give one or more vanishing quark masses, or degenerate quark masses, which is unphysical. That's why we shall not consider this model for our study anymore.
\subsection{$S_3$-symmetric 3HDM}

\subsubsection{Model details and Veltman coefficients}
This is yet another variant of the 3HDM that is symmetric under the discrete 
group $S_3$. The scalar potential is given by: 
\bea
V(\phi) &=& \mu_{11}^2 \left(\Phi_1^\dag \Phi_1 + \Phi_2^\dag \Phi_2\right) +
\mu_{33}^2 \Phi_3^\dag \Phi_3 + \lambda_1 \left(\Phi_1^\dag \Phi_1 + \Phi_2^\dag \Phi_2\right)^2 \nonumber \\
&& + \lambda_2 \left(\Phi_1^\dag \Phi_2 - \Phi_2^\dag \Phi_1\right)^2+ \lambda_3 \left[\left(\Phi_1^\dag \Phi_2 + \Phi_2^\dag \Phi_1\right)^2
+ \left(\Phi_1^\dag \Phi_1 - \Phi_2^\dag \Phi_2\right)^2\right] \nonumber\\
&& + \lambda_4 \left[\left(\Phi_3^\dag \Phi_1\right)\left(\Phi_1^\dag \Phi_2 + \Phi_2^\dag \Phi_1\right) + \left(\Phi_3^\dag \Phi_2\right)\left(\Phi_1^\dag \Phi_1 - \Phi_2^\dag \Phi_2\right) + \rm{h.c.}\right] \nonumber \\
&& + \lambda_5 \left(\Phi_3^\dag \Phi_3\right)\left(\Phi_1^\dag \Phi_1 + \Phi_2^\dag \Phi_2\right) + \lambda_6 \left[\left(\Phi_3^\dag \Phi_1\right)\left(\Phi_1^\dag \Phi_3\right) + \left(\Phi_3^\dag \Phi_2\right)\left(\Phi_2^\dag \Phi_3\right) \right] \nonumber \\
&& + \lambda_7  \left[\left(\Phi_3^\dag \Phi_1\right)\left(\Phi_3^\dag \Phi_1\right)+ \left(\Phi_3^\dag \Phi_2\right)\left(\Phi_3^\dag \Phi_2\right)+ \rm{h.c.} \right] + \lambda_8 \left(\Phi_3^\dag \Phi_3\right)^2 \,,
\label{s3pot}
\eea
We choose all the quartic couplings to be real to annul $CP-$violation coming
from the scalar sector. Similar to the $A_4$-symmetric case, the scalar potential in Eq.(\ref{s3pot}) permits multiple alignment of the VEVs 
$v_1, v_2, v_3$~\cite{Das:2014fea}. The ones we take up for our illustration are 
\besub
\bea
\text{case(a)}: v_1, v_2, v_3 \neq 0: v_1 = \sqrt{3} v_2 \label{s3active} \\ 
\text{case(b)}: v_1 = v_2 = 0; v_3 = 246 \text{GeV}\label{s3inert}.
\eea
\eesub
For case (a), tan$\beta = \frac{2 v_2}{v_3}$ is defined. 

Similar to a 2HDM,
the scalar mass matrices here are diagonalised by the action of the mixing angles $\alpha$ and $\beta$. The physical
spectrum consists of $h,H_{1,2},A_{1,2},H^\pm_{1,2}$, whose charge and $CP$-quantum numbers are same as in the $A_4$-symmetric case.
Of these, $h$ is identified with the 125 GeV Higgs. In addition one identifies $\alpha = \beta - \frac{\pi}{2}$ as the \emph{alignment limit}, analogously with the 2HDM, where couplings of $h$ with fermions and gauge bosons become SM-like.
The masses of the physical Higgses and the details of the diagonalisation
can be seen in~\cite{Das:2014fea} and thus, are not repeated here.  

Assuming that the first two fermion generations are $S_3$ doublets and the third generation is a singlet under the same, leads to the following Yukawa
Lagrangian, for the $u$-quarks:
\bea
\mathcal{L}_Y^{(u)} &=& -y_{1u} (\bar{Q}_1 \tilde{\Phi_3} u_{1R} + \bar{Q}_2 \tilde{\Phi_3} u_{2R}) - y_{2u} [(\bar{Q}_1 \tilde{\Phi_2}  + \bar{Q}_2 \tilde{\Phi_1} ) u_{1R} + (\bar{Q}_1 \tilde{\Phi_1}  - \bar{Q}_2 \tilde{\Phi_2} ) u_{2R}] \nonumber \\
&& - y_{3u} \bar{Q}_3 \tilde{\Phi_3} u_{3R} - y_{4u} \bar{Q}_3 (\tilde{\Phi_1} u_{1R} + \tilde{\Phi_2} u_{2R}) - y_{5u} (\bar{Q}_1 \tilde{\Phi_1} + \bar{Q}_2 \tilde{\Phi_2}) u_{3R} + { \rm h.c.} \,
\label{uYuk}
\eea
The corresponding Lagrangian for the $d$-quarks and the charged leptons
can be found by a straightforward replacement of the fermion indices and
by changing $\tilde{\Phi} \rightarrow \Phi$. 

In this limit, the flavour basis ($u_1,u_2,u_3$) is related to the mass eigenbasis ($u,c,t$) as
\bea
\begin{pmatrix}
u_1 \\
u_2 \\
u_3
\end{pmatrix} = \begin{pmatrix}
                \frac 1 2 & \frac{\sqrt{3}}{2} & 0 \\
                -\frac{\sqrt{3}}{2}  & \frac{1}{2} & 0 \\
                  0  & 0 & 1      
                \end{pmatrix} 
                \begin{pmatrix}
u \\
c\\
t
\end{pmatrix}
\eea
This allows us to determine the couplings between the quarks and the Higgs
bosons in their respective mass bases. We list the VCs corresponding
to the case (a) of Eq.(\ref{s3active}) below.
\besub
\bea
VC_{H_2} &=&  \frac{1}{8} (2 (37 \lambda_1 - 8 \lambda_2 + 19 \lambda_3 + 4 (2 \lambda_5 + \lambda_6 - 6 (y_{2u}^2 + y_{2d}^2)))+4
   \sqrt{3} \cos \beta  (\lambda_1+\lambda_3) \nonumber \\
&& + 6 \cos 2 \beta 
   (\lambda_3 - \lambda_1)-3 \lambda_4 \sin 2 \beta )
    + \frac{3}{4} g^2_1 + \frac{9}{4} g^2_2 \nonumber \\
VC_h &=& \frac{1}{16} (\cos 2 \alpha  (2 \sqrt{3} \cos \beta  (2 \lambda_1-2 \lambda_3-\lambda_5)+3 \cos 2 \beta  (\lambda_5-2 (\lambda_1+\lambda_3))  + \frac{3}{4} g^2_1 + \frac{9}{4} g^2_2 \nonumber \\
&& +74 \lambda_1-16 \lambda_2+26 \lambda_3-13 \lambda_5-8 (\lambda_6+6 (\lambda_8-2 y_{1u}^2-2 y_{1d}^2+2 y_{2u}^2+2 y_{2d}^2 \nonumber \\
&& -y_{3u}^2-y_{3d}^2))) + 2 \cos \beta  (\sqrt{3} (2 \lambda_1-2 \lambda_3+2 \lambda_4 \sin (2 \alpha )+\lambda_5)-3 \sin 2 \alpha  \sin \beta  (\lambda_6+2 \lambda_7)) \nonumber \\
&&   -3 \cos 2 \beta  (2 (\lambda_1+\lambda_3)+\lambda_5)+74 \lambda_1-16 \lambda_2+26
   \lambda_3+6 \lambda_4 \cos ^2\alpha  \sin 2 \beta  \nonumber \\
&&   +12 \lambda_4 \sin 2 \alpha  \cos ^2 \beta +45 \lambda_5+24 (\lambda_6+2 \lambda_8-4 y_{1u}^2-4 y_{1d}^2 \nonumber \\
&& -4 y_{2u}^2-4 y_{2d}^2-2 y_{3u}^2-2 y_{3d}^2)) \nonumber \\
VC_{H_1} &=& \frac{1}{16} \cos 2 \alpha  (2 \sqrt{3} \cos \beta  (-2 \lambda_1+2 \lambda_3+\lambda_5)+\cos 2 \beta  (6 (\lambda_1+\lambda_3)-3 \lambda_5) \nonumber \\
&& -74 \lambda_1 + 16 \lambda_2 - 26 \lambda_3-3 \lambda_4 \sin 2 \beta +13 \lambda_5 + 8 (\lambda_6+6 (\lambda_8-2 y_{1u}^2-2 y_{1d}^2+2 y_{2u}^2  \nonumber \\
&& + 2 y_{2d}^2- y_{3u}^2-y_{3d}^2)))+2 \sqrt{3} \cos (\beta ) (2 \lambda_1 -2 \lambda_3 + \lambda_5) \nonumber \\
&& -3 \cos 2 \beta  (2 (\lambda_1+\lambda_3)+\lambda_5)+74 \lambda_1-16 \lambda_2 + 26 \lambda_3 + 3 \lambda_4 \sin 2 \beta  \nonumber \\
&& +2 \sin 2 \alpha  \cos \beta  (3 \sin \beta  (\lambda_6 + 2 \lambda_7) \nonumber \\
&& -2 \lambda_4 (3 \cos \beta +\sqrt{3}))+45 \lambda_5 +24 (\lambda_6 + 2 \lambda_8 - 4  y_{1u} ^2-4  y_{1d}^2-4 y_{2u}^2 \nonumber \\
&& -4 y_{2d}^2-2 y_{3u}^2-2 y_{3d}^2))  + \frac{3}{4} g^2_1 + \frac{9}{4} g^2_2\nonumber \\
VC_{A_2} &=& \frac{1}{32}(8 \cos 2 \beta  (18 \lambda_1-4 \lambda_2 +6 \lambda_3 - 4 \lambda_5 
 -2 \lambda_6 -13 \lambda_8 + 12 (2 y_{1u}^2+2 y_{1d}^2 \nonumber \\
 && -2 y_{2u}^2-2
   y_{2d}^2 + y_{3u}^2 + y_{3d}^2)) 
+2 \sqrt{3} \cos \beta (6 \lambda_1 - 6 \lambda_3 + 4
   \lambda_4 \sin (2 \beta )+\lambda_5) \nonumber \\
&&   +2 \sqrt{3} \cos (3 \beta ) (2 \lambda_1-2 \lambda_3-\lambda_5)+\cos (4
   \beta ) (8 \lambda_1 + 8 \lambda_3 - 11 (\lambda_5 + \lambda_6 + 2 \lambda_7) \nonumber \\
&&   +14 \lambda_8)+136 \lambda_1-32 \lambda_2 +40 \lambda_3 + 10 \lambda_4 \sin (2 \beta ) \nonumber \\
&& -19 \lambda_4 \sin 4 \beta + 91 \lambda_5 + 43 \lambda_6-2
   (5 \lambda_7 - 45 \lambda_8 + 48 (2 y_{1u}^2+2 y_{1d}^2 + 2 y_{2u}^2 \nonumber \\
&&   + 2 y_{2d}^2+ y_{3u}^2 + y_{3d}^2)))  + \frac{3}{4} g^2_1 + \frac{9}{4} g^2_2 \nonumber \\
VC_{A_1} &=& \frac{1}{8} (2 (37 \lambda_1-8 \lambda_2 + 19 \lambda_3 + 4 (2 \lambda_5 + \lambda_6 -12
   (y_{2u}^2 + y_{2d}^2))) \nonumber \\
&&   + 4 \sqrt{3} \cos \beta  (\lambda_1 + \lambda_3)+6 \cos 2 \beta 
   (\lambda_3 - \lambda_1)-3 \lambda_4 \sin (2 \beta ))
    + \frac{3}{4} g^2_1 + \frac{9}{4} g^2_2 \nonumber \\
   \eea
   \bea
VC_{H_2^\pm} &=& \frac{1}{32}(4 \cos 2 \beta  (7 \lambda_1 + \lambda_2 +4
  \lambda_3 -14 \lambda_5 - 7 \lambda_6 - 24 (\lambda_8 - 2 y_{1u}^2 \nonumber \\
  && + 4(-y_{1d}^2 + y_{2u}^2 + y_{2d}^2)))+2 \sqrt{3} \cos \beta  (- 3 \lambda_1 + 3 \lambda_2 + 2 \lambda_4 \sin 2 \beta  + \lambda_5 +\lambda_6) \nonumber \\
  && -2 \sqrt{3} \cos 3 \beta (\lambda_1 - \lambda_2 + \lambda_5 + \lambda_6) + 3 \cos 4 \beta 
   (-\lambda_1 + \lambda_2 +2 (\lambda_5 + \lambda_6))+31 \lambda_1 \nonumber \\
   && + \lambda_2 + 2(8 \lambda_3 +3 (11 \lambda_5 + 5 \lambda_6 +16 (\lambda_8 -2(y_{1u}^2 + 2 (y_{1d}^2 + y_{2u}^2 + y_{2d}^2))))) \nonumber \\
   && +96 \lambda_4 \sin (\beta ) \cos ^3(\beta )+192 y_{1u} y_{2u} \sin (2 \beta ))  + \frac{3}{4} g^2_1 + \frac{9}{4} g^2_2 \nonumber \\
VC_{H_1^\pm} &=& \frac{1}{32} (2 \sqrt{3} \cos (\beta ) (29 \lambda_1 - 5 \lambda_2 + 8 \lambda_3 -30 \lambda_4 \sin (2 \beta )+17 \lambda_5 + 13 \lambda_6) \nonumber \\
&& +3 \cos 2 \beta  (-39 \lambda_1 + 43 \lambda_2 + 4 (\lambda_3 +4(\lambda_5 + \lambda_6)))+6 \sqrt{3} \cos 3 \beta  (-3 \lambda_1 + 3 \lambda_2 + \lambda_5 + \lambda_6 ) \nonumber \\
&& + 299 \lambda_1 -127 \lambda_2 + 44 \lambda_3 - 216 \lambda_4 \sin 2 \beta +16 (7 \lambda_5 + 5 \lambda_6 - 48 (y_{2u}^2 + y_{2d}^2))) \nonumber \\
&&
 + \frac{3}{4} g^2_1 + \frac{9}{4} g^2_2 \,. 
\eea
\eesub

The inert case of the $S_3$-symmetric 3HDM can describe DM phenomenology successfully when a $Z_2$-symmetry is imposed that demands $\Phi_{1,2}$ to carry negative $Z_2$ charges and all other fields to 
carry positive charges~\cite{Chakrabarty:2015kmt}. In this case therefore, fermions are forbidden to couple to $\Phi_{1,2}$.
Accordingly, the VCs for $h$ and $S = H_{1,2},A_{1,2},H^pm_{1,2}$ have the following forms:
\besub
\bea
VC_h &=& 4 \l_5 + 2\l_6 + 6\l_8 + \frac{3}{4}g_1^2 + \frac{9}{4}g_2^2
 - 6 y^2_t \label{VCh_inerts3}\\
VC_{S} &=& 10 \l_1 - 2 \l_2 + 4 \l_3 + 2 \l_5 + \l_6+ \frac{3}{4}g_1^2 + \frac{9}{4}g_2^2 \label{VCH_inerts3} 
\eea
\eesub

\subsubsection{Analysis and results}
The masses of the physical scalars and the mixing angles $\a$ and $\b$
are taken as the independent parameters in this part.
We choose $m_h = 125$ GeV and vary the rest of the parameters in the 
following ranges:
\besub
\bea
-0.4 \leq \cos(\b - \a) \leq 0.4, ~~0.1 < \text{tan}\beta < 50 \\
100 ~\text{GeV} < m_{A_{1,2}}, m_{H^\pm_{1,2}} < 1 ~\text{TeV},
~~m_h < m_{H_{1,2}} < 1 ~\text{TeV}
\eea
\eesub
We ensure the model remains perturbative by demanding $|\l_i| \leq 4 \pi$ 
throughout. The scalar potential must be bounded from below in all directions
in order to preserve the stability of the EW vacuum, and thus, appropriate stability conditions~\cite{Das:2014fea} were imposed. In addition, unitarity of gauge-boson scattering was obeyed by requiring that the eigenvalues of the $2 \rightarrow 2$ scalar scattering matrix do not exceed 8$\pi$. Much like the previous sections, $VC_h = 0$ was demanded for the entire parameter space. 

\begin{figure} 
\begin{center}
%
\includegraphics[scale=0.40]{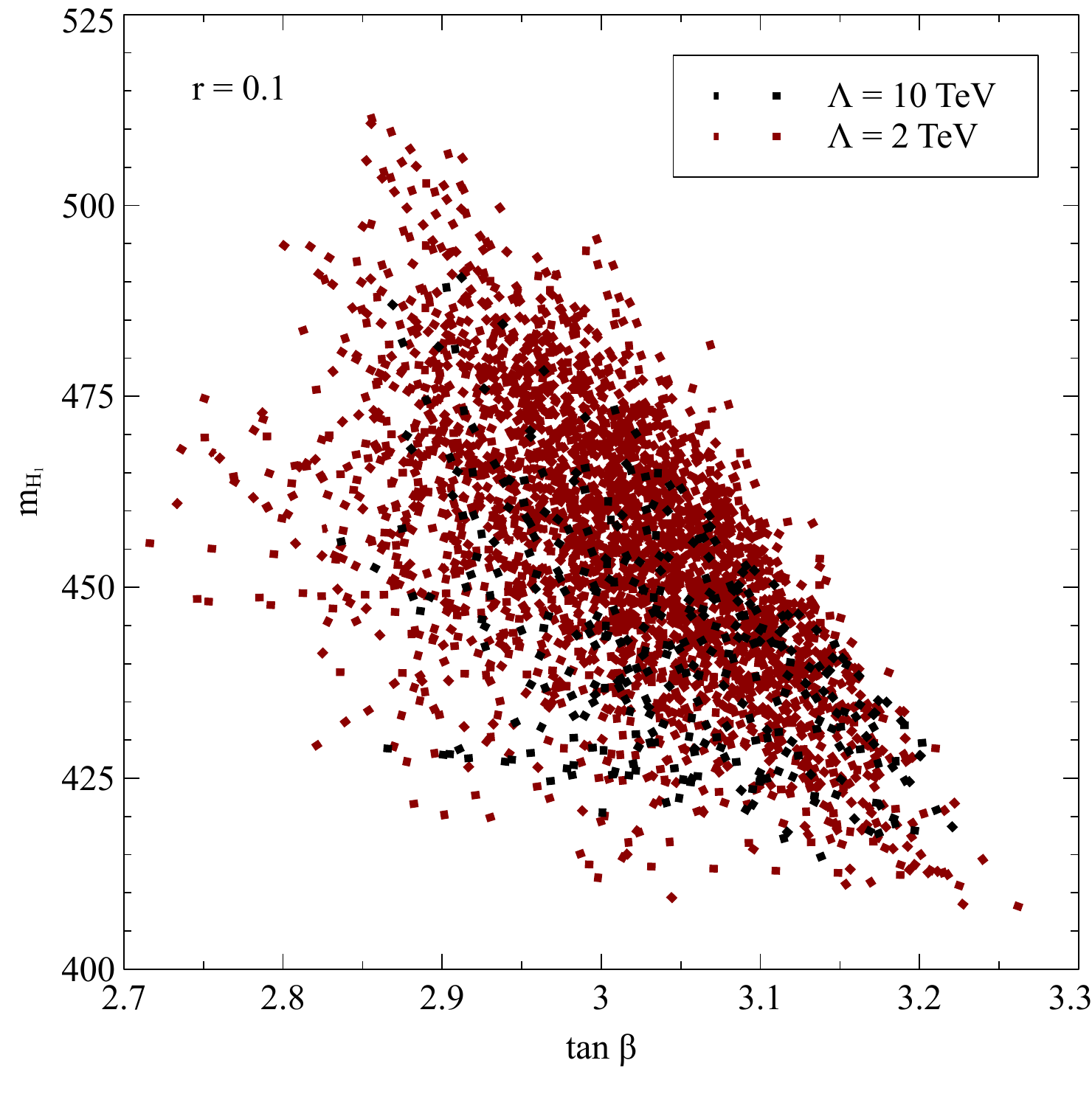}~~~ 
\includegraphics[scale=0.40]{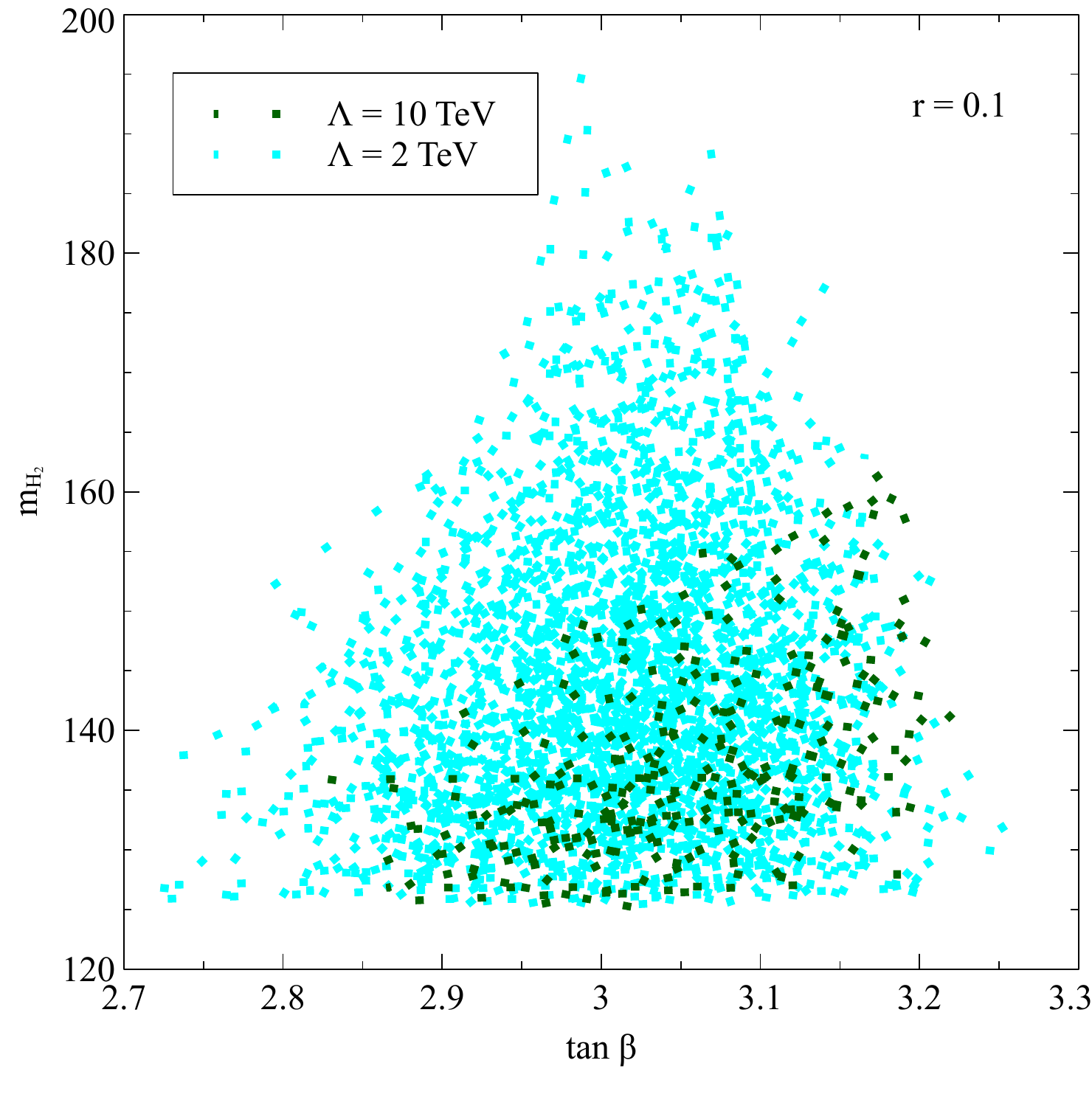} \\
\includegraphics[scale=0.40]{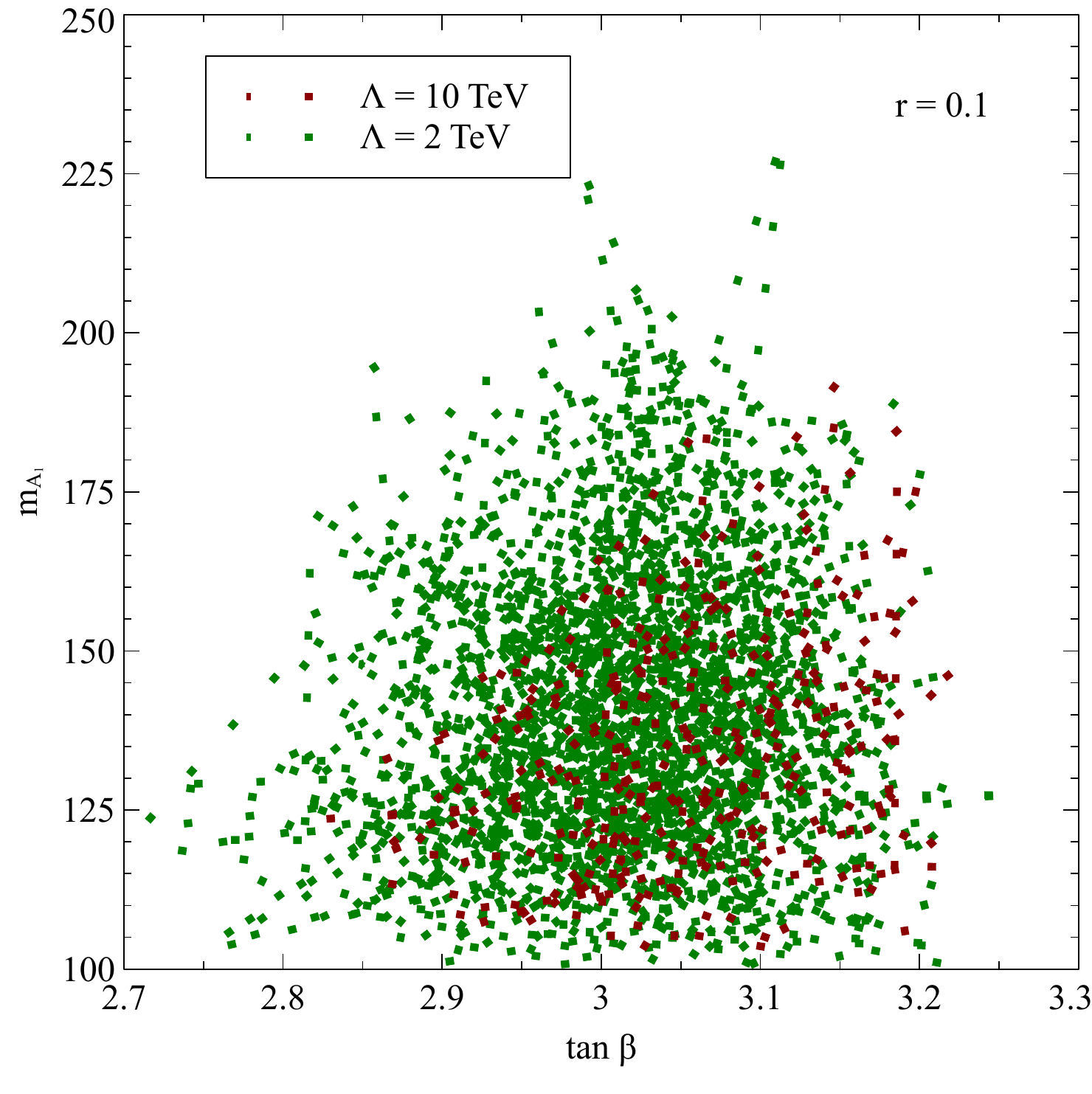}
\includegraphics[scale=0.40]{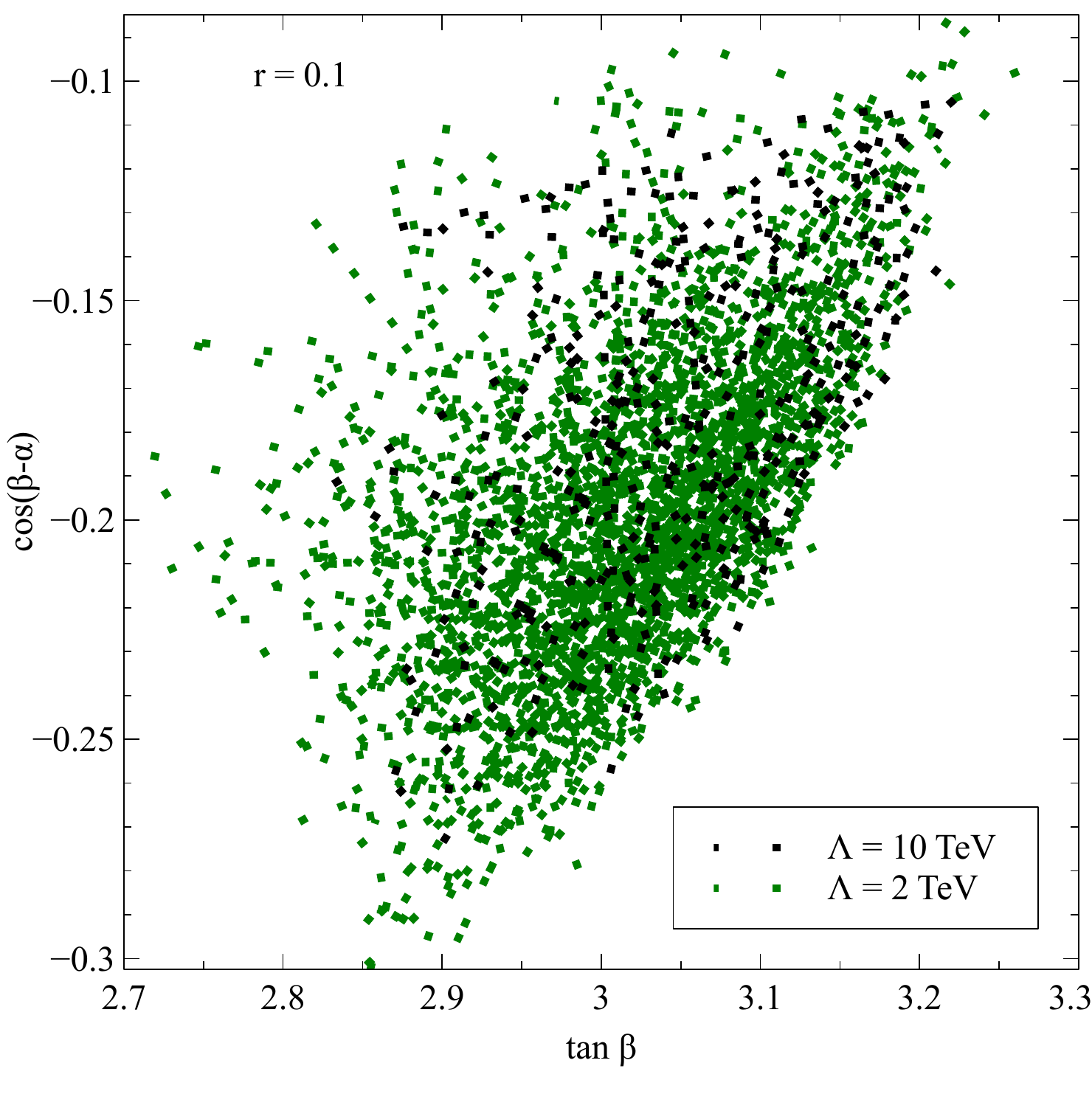}
\includegraphics[scale=0.40]{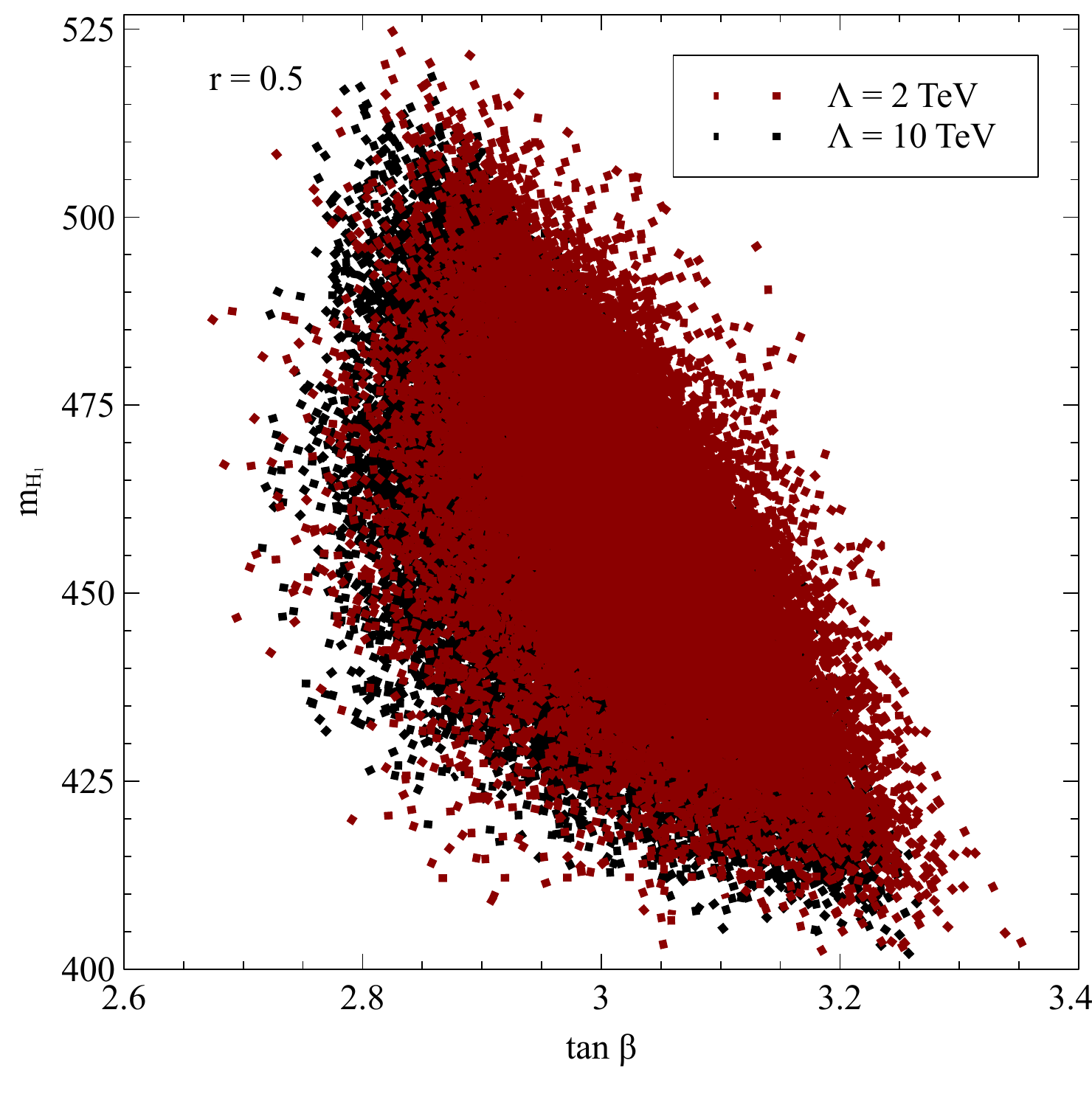}
\includegraphics[scale=0.40]{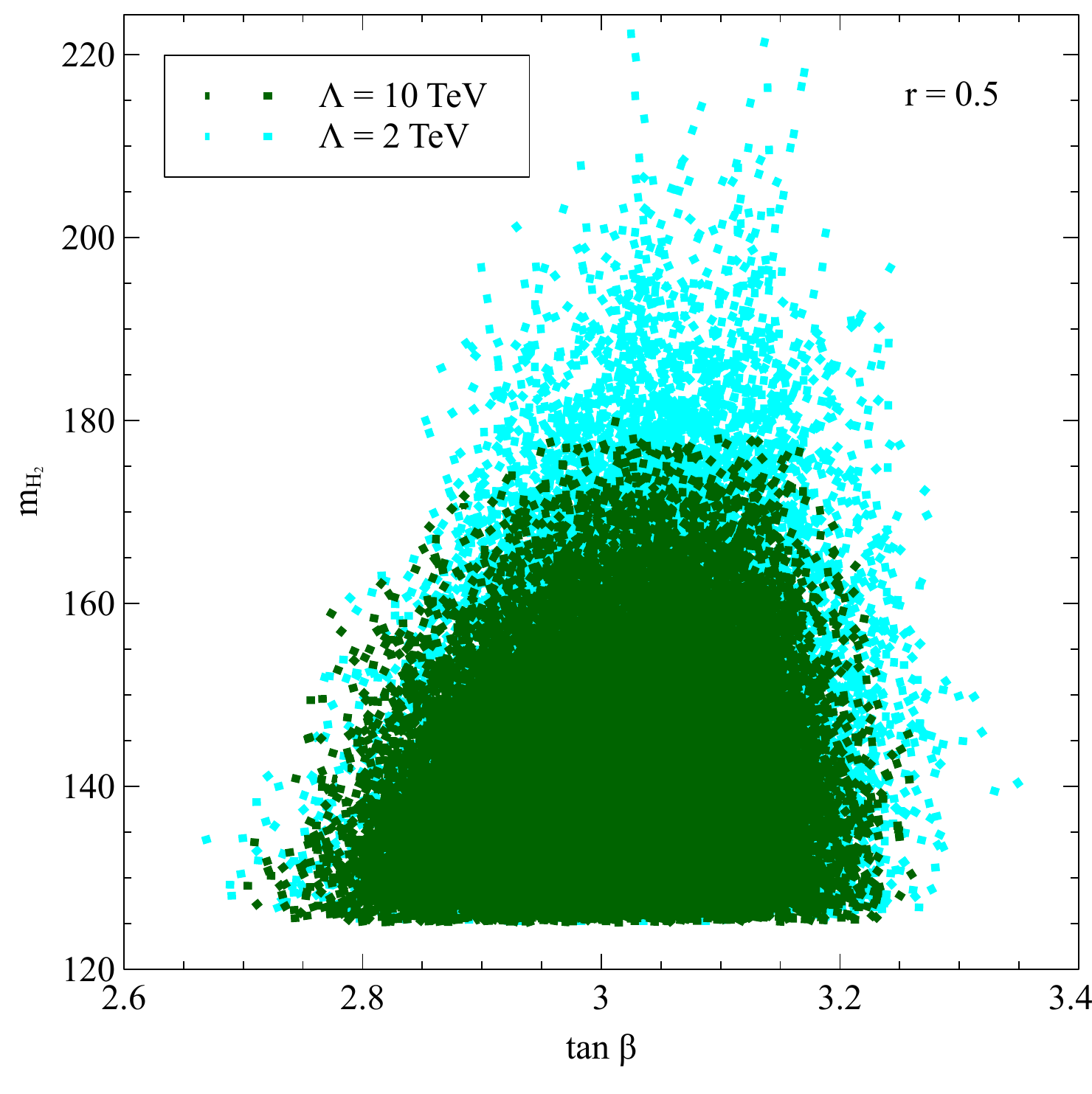}
\includegraphics[scale=0.40]{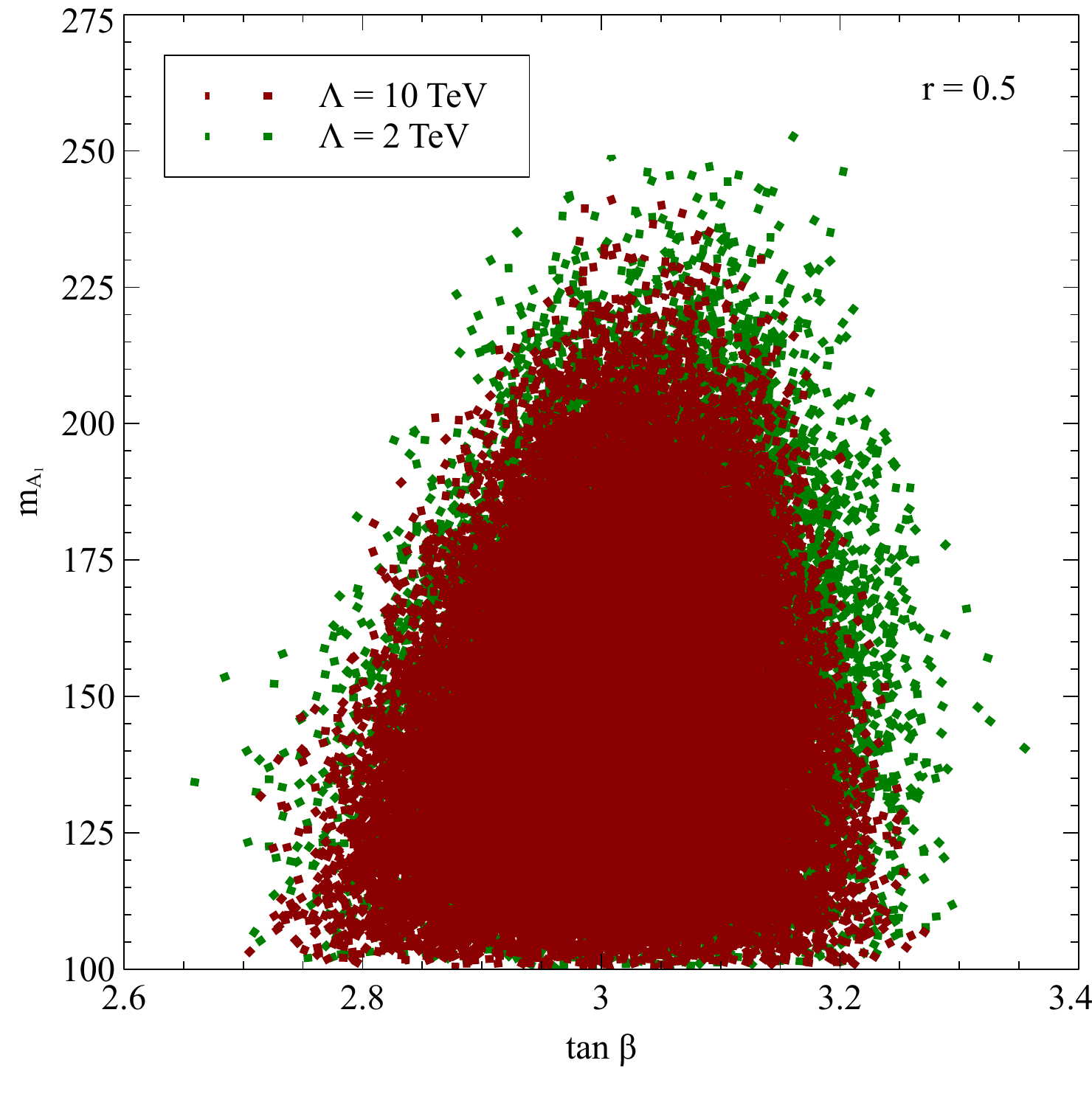}
\includegraphics[scale=0.40]{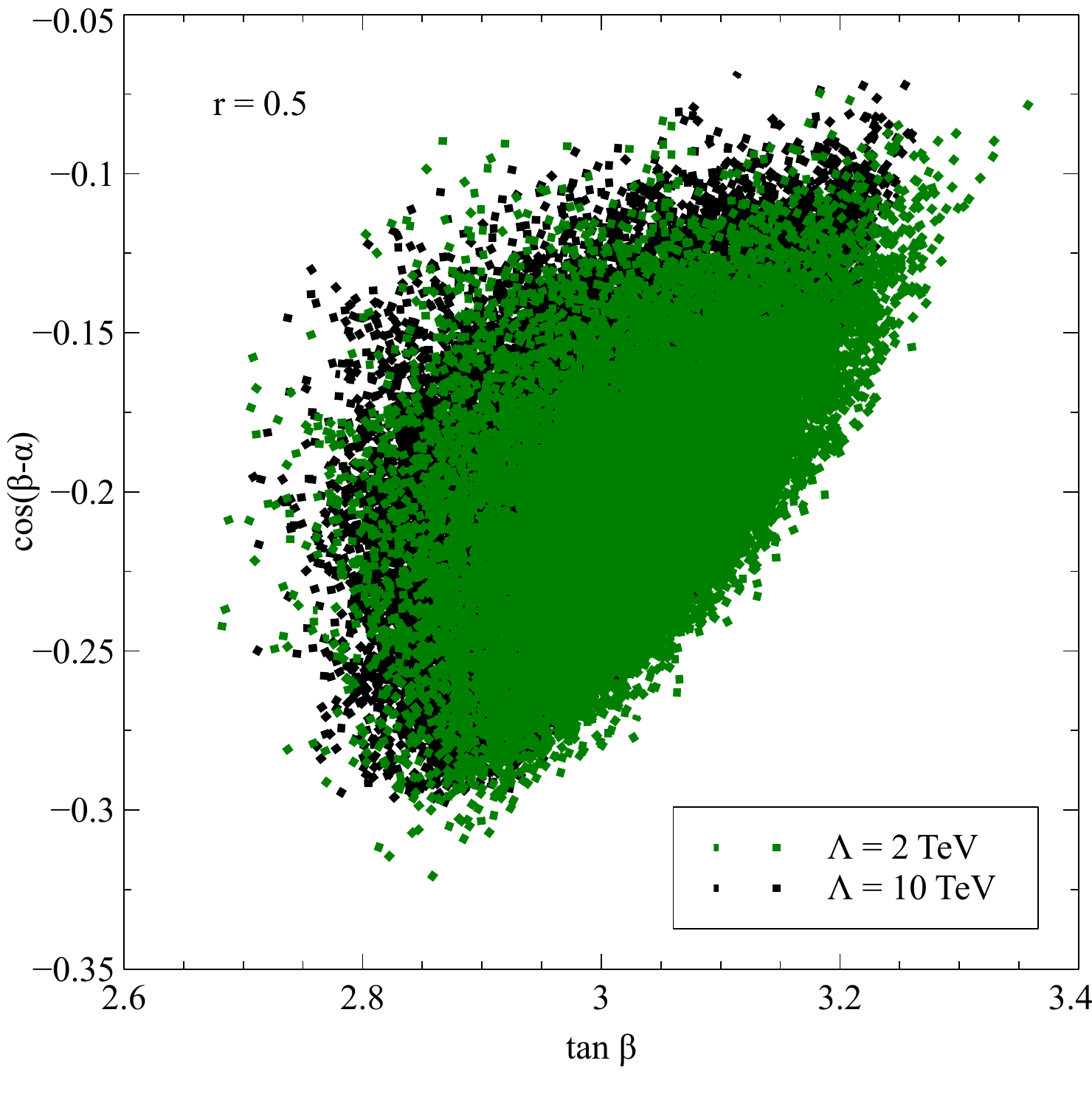}
\caption{Allowed parameter spaces in the tan$\beta$ vs scalar mass and
tan$\beta$ vs cos($\b-\a$) planes for $\L = 2,~10$ TeV and $r = 0.1, ~0.5$.
The colour coding is explained in the legends.}
\label{f:paramspace_s3}
\end{center}
\end{figure}

In the case where $v_1,v_2,v_3\neq 0$, we find it possible to fine-tune the
VCs corresponding to $H_1,H_2$ and $A_2$ to an $\mathcal{O}$(0.01). However,
the ones for the remaining scalars take larger values. More precisely, the 
parameter space leading to $VC_{H_{1,2}}, VC_{A_1} \simeq \mathcal{O}$(0.01)
gives $VC_{A_2} \simeq -50, VC_{H^\pm_1} \simeq -50$ and 
$VC_{H^\pm_2} \simeq 50$. Hence, a partial fine-tuning can be achieved in this model. However, demanding even such a partial fine-tuning can put stringent
constraints on the masses and the mixing angles. Similar to what was done
in case of the 2HDMs, we present the allowed ranges of the scalar masses
and cos$(\b-\a)$ for $\L$ = 2 TeV and 10 TeV and $r$ = 0.1 and 0.5,
in Fig.\ref{f:paramspace_s3}. 

The scan results shown in Fig.\ref{f:paramspace_s3} point to an important 
difference between the findings corresponding to a 2HDM and the $S_3$ symmetric 3HDM. That is, strict upper bounds on the masses appear in case of the latter. For instance, $H_1$ and $A_1$ cannot have masses exceeding 250 GeV. Besides, $H_2$ must have a mass in the [400 GeV, 520 GeV] range. This
particular feature of the spectrum is direct consequence of demanding small VCs for these scalars. Moreover, tan$\beta$ becomes $\sim$ 3, which in turn
restricts the quark couplings of the non-standard scalars. Altogether, a 
constrained mass spectrum and appropriately restricted quark couplings is
an attractive scenario to be probed at the colliders. For instance, production of the $S_3$ scalars and their decay to quarks can lead to interesting multi-jet signals that can verify or falsify this model.

\begin{figure} 
\begin{center}
%
\includegraphics[scale=0.38]{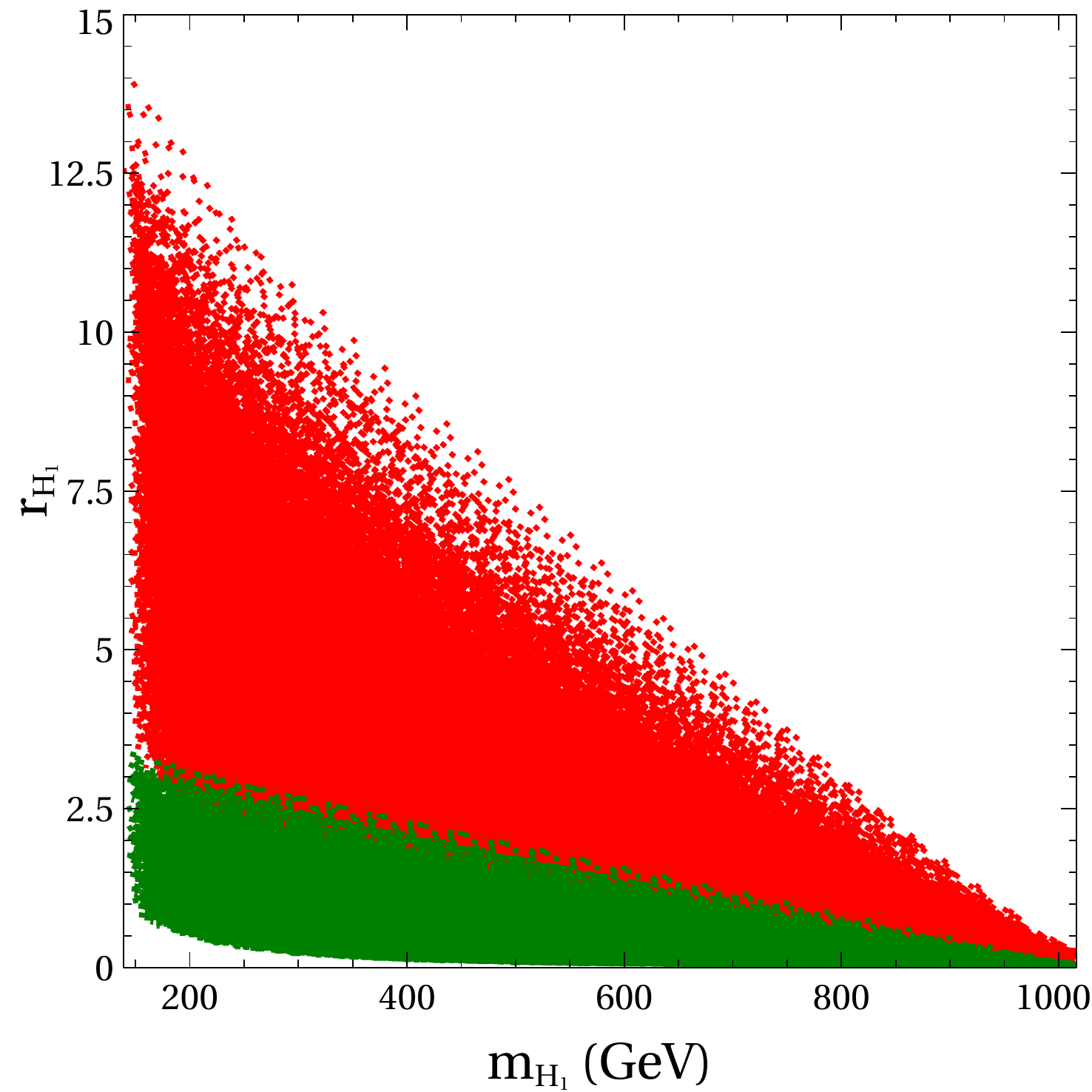}~~~ 
\caption{$r_H$ plotted versus $m_H$ for $\L = $ 2 TeV (green) and $\L = 10$ TeV (red).}
\label{f:rH-mH_s3}
\end{center}
\end{figure}

The results in case where $\Phi_2$ and $\Phi_3$ remain inert, qualitatively resemble those in
case of the IDM. This is again because the inert scalars do not couple to the fermions in this case. Large contributions to $\delta m^2_{S}$ are noted in this case, where, $S$ denotes the inert scalars. The lowest value of the corresponding VC is $\simeq$ 2.8. Therefore, this scenario is still 
somewhat better than the IDM quantitatively. Defining $r_S = \frac{\delta m^2_S}{m^2_S}$ as usual, we display the variation of $r_H$ versus $m_H$
in Fig.\ref{f:rH-mH_s3}.
The behaviour of the other $r_S$ are similar. The slight improvement with respect to the IDM noted here can be traced back to the negative sign of the coefficient of $\l_2$ in $VC_S$ (Eq.(\ref{VCh_inerts3})), which is obviously an
artifact of the $S_3$-symmetry.
\section{Summary and Conclusions}
\label{Summary}
The scalar mass not being protected by any symmetry, receives quadratically divergent correction term originating from the self-energy diagrams. Assuming the presence of some yet-to-be discovered symmetry protecting the scalar mass and no fine-tuning between the bare mass and corresponding radiative correction, one can set the quadratic divergence to be exactly equal to zero or keep it at some manageable level. It leads to the well known Veltman condition.

In this paper, the fine-tuning problem of Higgs mass has been addressed in light of various multi-Higgs doublet models, {\em i.e.} two Higgs doublet models, three Higgs doublet models etc. With the objective to cancel the quadratic divergent radiative correction to the scalar mass, we have tried to satisfy the Veltman conditions for all the physical scalars present in the model. The parameter space being compatible with several constraints like : stability conditions, unitarity, perturbativity, alignment limit etc, it has been found that it is not possible to satisfy the Veltman conditions for all the physical scalars simultaneously. Thus we aim to satisfy the Veltman condition exactly for the SM Higgs atleast and try to keep the quadratic divergences for the other scalars within manageable limit.

During the analysis with 2HDMs, we have used type-I, type-II, lepton specific, flipped, $S_3$-symmetric, inert doublet models for completeness of discussion. Similarly in case of 3HDM, the variants are $A_4$-symmetric, $S_4$-symmetric and $S_3$-symmetric 3HDMs. Following are our main observations :
\begin{itemize}
\item Out of the four canonical 2HDMs, the type II and flipped
models are most attractive from the perspective of fine-tuning. 
This leads to tight bounds on tan$\beta$ and cos($\b - \a)$ 
in these two cases. 

\item Since in IDM the physical scalars other than $h$ do not contain the fermionic contribution in their Veltman conditions, it is not possible to satisfy it only through cancellation among the terms present from the bosonic contribution alone. For the other scalars, we have tried to adjust the ratio $r_\Phi$ to make the correction as small as possible.
\item For $S_3$-symmetric 2HDM, Veltman condition for SM Higgs and Heavy Higgs can be satisfied exactly by adjusting the model parameters. Whereas for the other scalars,  smaller degree of fine-tuning (larger $r_\Phi$) is required for increasing $\Lambda$ to keep the correction tiny.                                                                                                                                                                                                                                                                                                                                                                                                                                                                                                                                            
\item In $A_4$-symmetric 3HDM, Veltman condition for $h, H_2(A_2), H_1^\pm(H_2^\pm)$ cannot be satisfied exactly due to the dominance of bosonic contribution over the fermionic one. On the other hand the Veltman conditions for $H_1$ and $A_1$ are far from being satisfied since $VC_{H_1}, VC_{A_1}$ do not contain any fermionic contribution to cancel the bosonic one. The Quadratic divergence of $h$ is kept in control by adjusting the minimum value of the ratio $r_h = 0.68$ for $\Lambda = 1$ TeV.
\item The $S_4$-symmetric 3HDM seems to be unfavoured due presence of one or more vanishing quark masses or degenerate quark masses. Therefore it is excluded from our discussion. 
\item {One observes partial fine-tuning in the masses of 
$S_3$-symmetric scalars when all the doublets receive VEVs. It becomes possible to keep the quadratic divergences in the masses of the $CP$-even neutral scalars and one of the $CP-$ odd scalars at a manageable level. Unlike the 2HDM, upper bounds on the scalar masses are obtained by virtue of fine-tuning. For the inert case, the quadratic divergences in the masses of the non-standard scalars cannot be controlled. However, this scenario fares slightly better than the IDM quantitatively.}
\end{itemize}

Therefore, we observe that an utility of these multi-Higgs doublet models lies in the very fact that the quadratic divergence in the mass of the observed scalar boson of mass 125 GeV can be controlled, or, exactly annulled. Adopting this top-down approach one can evade the fine-tuning problem of Higgs mass as well as the other physical scalar mass without the application of any specific symmetry. Whenever the Veltman condition cannot be satisfied, the radiative correction is kept under control by proper tuning of the bare mass and correction term, {\em i.e.} by adjusting $r_\Phi$, depending on the cutoff scale $\Lambda$. 
\section{Acknowledgement}
We thank Prof. Biswarup Mukhopadhyaya and Prof. Anirban Kundu for fruitful discussions. This work is partially supported by funding available from the Department of Atomic Energy, Government of India, 
for the Regional Center for Accelerator- based Particle Physics (RECAPP), Harish-Chandra Research Institute. NC also acknowledges the financial support
received from National Center for Theoretical Sciences, Hsinchu, Taiwan.
\bibliography{3HDM_ref}{}

\begin{thebibliography}{31}%
\makeatletter
\providecommand \@ifxundefined [1]{%
 \@ifx{#1\undefined}
}%
\providecommand \@ifnum [1]{%
 \ifnum #1\expandafter \@firstoftwo
 \else \expandafter \@secondoftwo
 \fi
}%
\providecommand \@ifx [1]{%
 \ifx #1\expandafter \@firstoftwo
 \else \expandafter \@secondoftwo
 \fi
}%
\providecommand \natexlab [1]{#1}%
\providecommand \enquote  [1]{``#1''}%
\providecommand \bibnamefont  [1]{#1}%
\providecommand \bibfnamefont [1]{#1}%
\providecommand \citenamefont [1]{#1}%
\providecommand \href@noop [0]{\@secondoftwo}%
\providecommand \href [0]{\begingroup \@sanitize@url \@href}%
\providecommand \@href[1]{\@@startlink{#1}\@@href}%
\providecommand \@@href[1]{\endgroup#1\@@endlink}%
\providecommand \@sanitize@url [0]{\catcode `\\12\catcode `\$12\catcode
  `\&12\catcode `\#12\catcode `\^12\catcode `\_12\catcode `\%12\relax}%
\providecommand \@@startlink[1]{}%
\providecommand \@@endlink[0]{}%
\providecommand \url  [0]{\begingroup\@sanitize@url \@url }%
\providecommand \@url [1]{\endgroup\@href {#1}{\urlprefix }}%
\providecommand \urlprefix  [0]{URL }%
\providecommand \Eprint [0]{\href }%
\providecommand \doibase [0]{http://dx.doi.org/}%
\providecommand \selectlanguage [0]{\@gobble}%
\providecommand \bibinfo  [0]{\@secondoftwo}%
\providecommand \bibfield  [0]{\@secondoftwo}%
\providecommand \translation [1]{[#1]}%
\providecommand \BibitemOpen [0]{}%
\providecommand \bibitemStop [0]{}%
\providecommand \bibitemNoStop [0]{.\EOS\space}%
\providecommand \EOS [0]{\spacefactor3000\relax}%
\providecommand \BibitemShut  [1]{\csname bibitem#1\endcsname}%
\let\auto@bib@innerbib\@empty
\bibitem [{\citenamefont {Veltman}(1981)}]{Veltman:1980mj}%
  \BibitemOpen
  \bibfield  {author} {\bibinfo {author} {\bibfnamefont {M.~J.~G.}\
  \bibnamefont {Veltman}},\ }\href@noop {} {\bibfield  {journal} {\bibinfo
  {journal} {Acta Phys. Polon.}\ }\textbf {\bibinfo {volume} {B12}},\ \bibinfo
  {pages} {437} (\bibinfo {year} {1981})}\BibitemShut {NoStop}%
\bibitem [{\citenamefont {Kundu}\ and\ \citenamefont
  {Raychaudhuri}(1996)}]{Kundu:1994bs}%
  \BibitemOpen
  \bibfield  {author} {\bibinfo {author} {\bibfnamefont {A.}~\bibnamefont
  {Kundu}}\ and\ \bibinfo {author} {\bibfnamefont {S.}~\bibnamefont
  {Raychaudhuri}},\ }\href {\doibase 10.1103/PhysRevD.53.4042} {\bibfield
  {journal} {\bibinfo  {journal} {Phys. Rev.}\ }\textbf {\bibinfo {volume}
  {D53}},\ \bibinfo {pages} {4042} (\bibinfo {year} {1996})},\ \Eprint
  {http://arxiv.org/abs/hep-ph/9410291} {arXiv:hep-ph/9410291 [hep-ph]}
  \BibitemShut {NoStop}%
\bibitem [{\citenamefont {Chakraborty}\ and\ \citenamefont
  {Kundu}(2013)}]{Chakraborty:2012rb}%
  \BibitemOpen
  \bibfield  {author} {\bibinfo {author} {\bibfnamefont {I.}~\bibnamefont
  {Chakraborty}}\ and\ \bibinfo {author} {\bibfnamefont {A.}~\bibnamefont
  {Kundu}},\ }\href {\doibase 10.1103/PhysRevD.87.055015} {\bibfield  {journal}
  {\bibinfo  {journal} {Phys. Rev.}\ }\textbf {\bibinfo {volume} {D87}},\
  \bibinfo {pages} {055015} (\bibinfo {year} {2013})},\ \Eprint
  {http://arxiv.org/abs/1212.0394} {arXiv:1212.0394 [hep-ph]} \BibitemShut
  {NoStop}%
\bibitem [{\citenamefont {Karahan}\ and\ \citenamefont
  {Korutlu}(2014)}]{Karahan:2014ola}%
  \BibitemOpen
  \bibfield  {author} {\bibinfo {author} {\bibfnamefont {C.~N.}\ \bibnamefont
  {Karahan}}\ and\ \bibinfo {author} {\bibfnamefont {B.}~\bibnamefont
  {Korutlu}},\ }\href {\doibase 10.1016/j.physletb.2014.03.063} {\bibfield
  {journal} {\bibinfo  {journal} {Phys. Lett.}\ }\textbf {\bibinfo {volume}
  {B732}},\ \bibinfo {pages} {320} (\bibinfo {year} {2014})},\ \Eprint
  {http://arxiv.org/abs/1404.0175} {arXiv:1404.0175 [hep-ph]} \BibitemShut
  {NoStop}%
\bibitem [{\citenamefont {Chakraborty}\ and\ \citenamefont
  {Kundu}(2014{\natexlab{a}})}]{Chakraborty:2014oma}%
  \BibitemOpen
  \bibfield  {author} {\bibinfo {author} {\bibfnamefont {I.}~\bibnamefont
  {Chakraborty}}\ and\ \bibinfo {author} {\bibfnamefont {A.}~\bibnamefont
  {Kundu}},\ }\href {\doibase 10.1103/PhysRevD.90.115017} {\bibfield  {journal}
  {\bibinfo  {journal} {Phys. Rev.}\ }\textbf {\bibinfo {volume} {D90}},\
  \bibinfo {pages} {115017} (\bibinfo {year} {2014}{\natexlab{a}})},\ \Eprint
  {http://arxiv.org/abs/1404.3038} {arXiv:1404.3038 [hep-ph]} \BibitemShut
  {NoStop}%
\bibitem [{\citenamefont {Biswas}\ and\ \citenamefont
  {Lahiri}(2016)}]{Biswas:2016bth}%
  \BibitemOpen
  \bibfield  {author} {\bibinfo {author} {\bibfnamefont {A.}~\bibnamefont
  {Biswas}}\ and\ \bibinfo {author} {\bibfnamefont {A.}~\bibnamefont
  {Lahiri}},\ }\bibfield  {booktitle} {\emph {\bibinfo {booktitle}
  {{Proceedings, 38th International Conference on High Energy Physics (ICHEP
  2016): Chicago, IL, USA, August 3-10, 2016}}},\ }\href@noop {} {\bibfield
  {journal} {\bibinfo  {journal} {PoS}\ }\textbf {\bibinfo {volume}
  {ICHEP2016}},\ \bibinfo {pages} {710} (\bibinfo {year} {2016})}\BibitemShut
  {NoStop}%
\bibitem [{\citenamefont {Chakraborty}\ and\ \citenamefont
  {Kundu}(2014{\natexlab{b}})}]{Chakraborty:2014xqa}%
  \BibitemOpen
  \bibfield  {author} {\bibinfo {author} {\bibfnamefont {I.}~\bibnamefont
  {Chakraborty}}\ and\ \bibinfo {author} {\bibfnamefont {A.}~\bibnamefont
  {Kundu}},\ }\href {\doibase 10.1103/PhysRevD.89.095032} {\bibfield  {journal}
  {\bibinfo  {journal} {Phys. Rev.}\ }\textbf {\bibinfo {volume} {D89}},\
  \bibinfo {pages} {095032} (\bibinfo {year} {2014}{\natexlab{b}})},\ \Eprint
  {http://arxiv.org/abs/1404.1723} {arXiv:1404.1723 [hep-ph]} \BibitemShut
  {NoStop}%
\bibitem [{\citenamefont {Antipin}\ \emph {et~al.}(2014)\citenamefont
  {Antipin}, \citenamefont {Mojaza},\ and\ \citenamefont
  {Sannino}}]{Antipin:2013exa}%
  \BibitemOpen
  \bibfield  {author} {\bibinfo {author} {\bibfnamefont {O.}~\bibnamefont
  {Antipin}}, \bibinfo {author} {\bibfnamefont {M.}~\bibnamefont {Mojaza}}, \
  and\ \bibinfo {author} {\bibfnamefont {F.}~\bibnamefont {Sannino}},\ }\href
  {\doibase 10.1103/PhysRevD.89.085015} {\bibfield  {journal} {\bibinfo
  {journal} {Phys. Rev.}\ }\textbf {\bibinfo {volume} {D89}},\ \bibinfo {pages}
  {085015} (\bibinfo {year} {2014})},\ \Eprint {http://arxiv.org/abs/1310.0957}
  {arXiv:1310.0957 [hep-ph]} \BibitemShut {NoStop}%
\bibitem [{\citenamefont {Masina}\ and\ \citenamefont
  {Quiros}(2013)}]{Masina:2013wja}%
  \BibitemOpen
  \bibfield  {author} {\bibinfo {author} {\bibfnamefont {I.}~\bibnamefont
  {Masina}}\ and\ \bibinfo {author} {\bibfnamefont {M.}~\bibnamefont
  {Quiros}},\ }\href {\doibase 10.1103/PhysRevD.88.093003} {\bibfield
  {journal} {\bibinfo  {journal} {Phys. Rev.}\ }\textbf {\bibinfo {volume}
  {D88}},\ \bibinfo {pages} {093003} (\bibinfo {year} {2013})},\ \Eprint
  {http://arxiv.org/abs/1308.1242} {arXiv:1308.1242 [hep-ph]} \BibitemShut
  {NoStop}%
\bibitem [{\citenamefont {Drozd}(2010)}]{Drozd:2012is}%
  \BibitemOpen
  \bibfield  {author} {\bibinfo {author} {\bibfnamefont {A.}~\bibnamefont
  {Drozd}},\ }\emph {\bibinfo {title} {{RGE and the Fine-Tuning Problem}}},\
  \href {http://inspirehep.net/record/1087035/files/arXiv:1202.0195.pdf}
  {Master's thesis},\ \bibinfo  {school} {Warsaw U.} (\bibinfo {year} {2010}),\
  \Eprint {http://arxiv.org/abs/1202.0195} {arXiv:1202.0195 [hep-ph]}
  \BibitemShut {NoStop}%
\bibitem [{\citenamefont {Pivovarov}\ and\ \citenamefont
  {Kim}(2008)}]{Pivovarov:2007dj}%
  \BibitemOpen
  \bibfield  {author} {\bibinfo {author} {\bibfnamefont {G.~B.}\ \bibnamefont
  {Pivovarov}}\ and\ \bibinfo {author} {\bibfnamefont {V.~T.}\ \bibnamefont
  {Kim}},\ }\href {\doibase 10.1103/PhysRevD.78.016001} {\bibfield  {journal}
  {\bibinfo  {journal} {Phys. Rev.}\ }\textbf {\bibinfo {volume} {D78}},\
  \bibinfo {pages} {016001} (\bibinfo {year} {2008})},\ \Eprint
  {http://arxiv.org/abs/0712.0402} {arXiv:0712.0402 [hep-ph]} \BibitemShut
  {NoStop}%
\bibitem [{\citenamefont {Branco}\ \emph {et~al.}(2012)\citenamefont {Branco},
  \citenamefont {Ferreira}, \citenamefont {Lavoura}, \citenamefont {Rebelo},
  \citenamefont {Sher},\ and\ \citenamefont {Silva}}]{Branco:2011iw}%
  \BibitemOpen
  \bibfield  {author} {\bibinfo {author} {\bibfnamefont {G.~C.}\ \bibnamefont
  {Branco}}, \bibinfo {author} {\bibfnamefont {P.~M.}\ \bibnamefont
  {Ferreira}}, \bibinfo {author} {\bibfnamefont {L.}~\bibnamefont {Lavoura}},
  \bibinfo {author} {\bibfnamefont {M.~N.}\ \bibnamefont {Rebelo}}, \bibinfo
  {author} {\bibfnamefont {M.}~\bibnamefont {Sher}}, \ and\ \bibinfo {author}
  {\bibfnamefont {J.~P.}\ \bibnamefont {Silva}},\ }\href {\doibase
  10.1016/j.physrep.2012.02.002} {\bibfield  {journal} {\bibinfo  {journal}
  {Phys. Rept.}\ }\textbf {\bibinfo {volume} {516}},\ \bibinfo {pages} {1}
  (\bibinfo {year} {2012})},\ \Eprint {http://arxiv.org/abs/1106.0034}
  {arXiv:1106.0034 [hep-ph]} \BibitemShut {NoStop}%
\bibitem [{\citenamefont {Bhattacharyya}\ and\ \citenamefont
  {Das}(2016)}]{Bhattacharyya:2015nca}%
  \BibitemOpen
  \bibfield  {author} {\bibinfo {author} {\bibfnamefont {G.}~\bibnamefont
  {Bhattacharyya}}\ and\ \bibinfo {author} {\bibfnamefont {D.}~\bibnamefont
  {Das}},\ }\href {\doibase 10.1007/s12043-016-1252-4} {\bibfield  {journal}
  {\bibinfo  {journal} {Pramana}\ }\textbf {\bibinfo {volume} {87}},\ \bibinfo
  {pages} {40} (\bibinfo {year} {2016})},\ \Eprint
  {http://arxiv.org/abs/1507.06424} {arXiv:1507.06424 [hep-ph]} \BibitemShut
  {NoStop}%
\bibitem [{\citenamefont {González~Felipe}\ \emph {et~al.}(2013)\citenamefont
  {González~Felipe}, \citenamefont {Serôdio},\ and\ \citenamefont
  {Silva}}]{Felipe:2013ie}%
  \BibitemOpen
  \bibfield  {author} {\bibinfo {author} {\bibfnamefont {R.}~\bibnamefont
  {González~Felipe}}, \bibinfo {author} {\bibfnamefont {H.}~\bibnamefont
  {Serôdio}}, \ and\ \bibinfo {author} {\bibfnamefont {J.~P.}\ \bibnamefont
  {Silva}},\ }\href {\doibase 10.1103/PhysRevD.87.055010} {\bibfield  {journal}
  {\bibinfo  {journal} {Phys. Rev.}\ }\textbf {\bibinfo {volume} {D87}},\
  \bibinfo {pages} {055010} (\bibinfo {year} {2013})},\ \Eprint
  {http://arxiv.org/abs/1302.0861} {arXiv:1302.0861 [hep-ph]} \BibitemShut
  {NoStop}%
\bibitem [{\citenamefont {Machado}\ \emph {et~al.}(2011)\citenamefont
  {Machado}, \citenamefont {Montero},\ and\ \citenamefont
  {Pleitez}}]{Machado:2010uc}%
  \BibitemOpen
  \bibfield  {author} {\bibinfo {author} {\bibfnamefont {A.~C.~B.}\
  \bibnamefont {Machado}}, \bibinfo {author} {\bibfnamefont {J.~C.}\
  \bibnamefont {Montero}}, \ and\ \bibinfo {author} {\bibfnamefont
  {V.}~\bibnamefont {Pleitez}},\ }\href {\doibase
  10.1016/j.physletb.2011.02.015} {\bibfield  {journal} {\bibinfo  {journal}
  {Phys. Lett.}\ }\textbf {\bibinfo {volume} {B697}},\ \bibinfo {pages} {318}
  (\bibinfo {year} {2011})},\ \Eprint {http://arxiv.org/abs/1011.5855}
  {arXiv:1011.5855 [hep-ph]} \BibitemShut {NoStop}%
\bibitem [{\citenamefont {Ivanov}\ and\ \citenamefont
  {Nishi}(2015)}]{Ivanov:2014doa}%
  \BibitemOpen
  \bibfield  {author} {\bibinfo {author} {\bibfnamefont {I.~P.}\ \bibnamefont
  {Ivanov}}\ and\ \bibinfo {author} {\bibfnamefont {C.~C.}\ \bibnamefont
  {Nishi}},\ }\href {\doibase 10.1007/JHEP01(2015)021} {\bibfield  {journal}
  {\bibinfo  {journal} {JHEP}\ }\textbf {\bibinfo {volume} {01}},\ \bibinfo
  {pages} {021} (\bibinfo {year} {2015})},\ \Eprint
  {http://arxiv.org/abs/1410.6139} {arXiv:1410.6139 [hep-ph]} \BibitemShut
  {NoStop}%
\bibitem [{\citenamefont {Ivanov}\ and\ \citenamefont
  {Vdovin}(2013)}]{Ivanov:2012fp}%
  \BibitemOpen
  \bibfield  {author} {\bibinfo {author} {\bibfnamefont {I.~P.}\ \bibnamefont
  {Ivanov}}\ and\ \bibinfo {author} {\bibfnamefont {E.}~\bibnamefont
  {Vdovin}},\ }\href {\doibase 10.1140/epjc/s10052-013-2309-x} {\bibfield
  {journal} {\bibinfo  {journal} {Eur. Phys. J.}\ }\textbf {\bibinfo {volume}
  {C73}},\ \bibinfo {pages} {2309} (\bibinfo {year} {2013})},\ \Eprint
  {http://arxiv.org/abs/1210.6553} {arXiv:1210.6553 [hep-ph]} \BibitemShut
  {NoStop}%
\bibitem [{\citenamefont {Degee}\ \emph {et~al.}(2013)\citenamefont {Degee},
  \citenamefont {Ivanov},\ and\ \citenamefont {Keus}}]{Degee:2012sk}%
  \BibitemOpen
  \bibfield  {author} {\bibinfo {author} {\bibfnamefont {A.}~\bibnamefont
  {Degee}}, \bibinfo {author} {\bibfnamefont {I.~P.}\ \bibnamefont {Ivanov}}, \
  and\ \bibinfo {author} {\bibfnamefont {V.}~\bibnamefont {Keus}},\ }\href
  {\doibase 10.1007/JHEP02(2013)125} {\bibfield  {journal} {\bibinfo  {journal}
  {JHEP}\ }\textbf {\bibinfo {volume} {02}},\ \bibinfo {pages} {125} (\bibinfo
  {year} {2013})},\ \Eprint {http://arxiv.org/abs/1211.4989} {arXiv:1211.4989
  [hep-ph]} \BibitemShut {NoStop}%
\bibitem [{\citenamefont {Das}\ \emph {et~al.}(2017)\citenamefont {Das},
  \citenamefont {Dey},\ and\ \citenamefont {Pal}}]{Das:2017zrm}%
  \BibitemOpen
  \bibfield  {author} {\bibinfo {author} {\bibfnamefont {D.}~\bibnamefont
  {Das}}, \bibinfo {author} {\bibfnamefont {U.~K.}\ \bibnamefont {Dey}}, \ and\
  \bibinfo {author} {\bibfnamefont {P.~B.}\ \bibnamefont {Pal}},\ }\href
  {\doibase 10.1103/PhysRevD.96.031701} {\bibfield  {journal} {\bibinfo
  {journal} {Phys. Rev.}\ }\textbf {\bibinfo {volume} {D96}},\ \bibinfo {pages}
  {031701} (\bibinfo {year} {2017})},\ \Eprint
  {http://arxiv.org/abs/1705.07784} {arXiv:1705.07784 [hep-ph]} \BibitemShut
  {NoStop}%
\bibitem [{\citenamefont {Chakrabarty}(2016)}]{Chakrabarty:2015kmt}%
  \BibitemOpen
  \bibfield  {author} {\bibinfo {author} {\bibfnamefont {N.}~\bibnamefont
  {Chakrabarty}},\ }\href {\doibase 10.1103/PhysRevD.93.075025} {\bibfield
  {journal} {\bibinfo  {journal} {Phys. Rev.}\ }\textbf {\bibinfo {volume}
  {D93}},\ \bibinfo {pages} {075025} (\bibinfo {year} {2016})},\ \Eprint
  {http://arxiv.org/abs/1511.08137} {arXiv:1511.08137 [hep-ph]} \BibitemShut
  {NoStop}%
\bibitem [{\citenamefont {Das}\ and\ \citenamefont {Dey}(2014)}]{Das:2014fea}%
  \BibitemOpen
  \bibfield  {author} {\bibinfo {author} {\bibfnamefont {D.}~\bibnamefont
  {Das}}\ and\ \bibinfo {author} {\bibfnamefont {U.~K.}\ \bibnamefont {Dey}},\
  }\href {\doibase 10.1103/PhysRevD.91.039905, 10.1103/PhysRevD.89.095025}
  {\bibfield  {journal} {\bibinfo  {journal} {Phys. Rev.}\ }\textbf {\bibinfo
  {volume} {D89}},\ \bibinfo {pages} {095025} (\bibinfo {year} {2014})},\
  \bibinfo {note} {[Erratum: Phys. Rev.D91,no.3,039905(2015)]},\ \Eprint
  {http://arxiv.org/abs/1404.2491} {arXiv:1404.2491 [hep-ph]} \BibitemShut
  {NoStop}%
\bibitem [{\citenamefont {Glashow}\ and\ \citenamefont
  {Weinberg}(1977)}]{Glashow:1976nt}%
  \BibitemOpen
  \bibfield  {author} {\bibinfo {author} {\bibfnamefont {S.~L.}\ \bibnamefont
  {Glashow}}\ and\ \bibinfo {author} {\bibfnamefont {S.}~\bibnamefont
  {Weinberg}},\ }\href {\doibase 10.1103/PhysRevD.15.1958} {\bibfield
  {journal} {\bibinfo  {journal} {Phys. Rev.}\ }\textbf {\bibinfo {volume}
  {D15}},\ \bibinfo {pages} {1958} (\bibinfo {year} {1977})}\BibitemShut
  {NoStop}%
\bibitem [{\citenamefont {Paschos}(1977)}]{Paschos:1976ay}%
  \BibitemOpen
  \bibfield  {author} {\bibinfo {author} {\bibfnamefont {E.~A.}\ \bibnamefont
  {Paschos}},\ }\href {\doibase 10.1103/PhysRevD.15.1966} {\bibfield  {journal}
  {\bibinfo  {journal} {Phys. Rev.}\ }\textbf {\bibinfo {volume} {D15}},\
  \bibinfo {pages} {1966} (\bibinfo {year} {1977})}\BibitemShut {NoStop}%
\bibitem [{\citenamefont {Barroso}\ \emph {et~al.}(2013)\citenamefont
  {Barroso}, \citenamefont {Ferreira}, \citenamefont {Ivanov},\ and\
  \citenamefont {Santos}}]{Barroso:2013awa}%
  \BibitemOpen
  \bibfield  {author} {\bibinfo {author} {\bibfnamefont {A.}~\bibnamefont
  {Barroso}}, \bibinfo {author} {\bibfnamefont {P.~M.}\ \bibnamefont
  {Ferreira}}, \bibinfo {author} {\bibfnamefont {I.~P.}\ \bibnamefont
  {Ivanov}}, \ and\ \bibinfo {author} {\bibfnamefont {R.}~\bibnamefont
  {Santos}},\ }\href {\doibase 10.1007/JHEP06(2013)045} {\bibfield  {journal}
  {\bibinfo  {journal} {JHEP}\ }\textbf {\bibinfo {volume} {06}},\ \bibinfo
  {pages} {045} (\bibinfo {year} {2013})},\ \Eprint
  {http://arxiv.org/abs/1303.5098} {arXiv:1303.5098 [hep-ph]} \BibitemShut
  {NoStop}%
\bibitem [{\citenamefont {Chakraborty}\ and\ \citenamefont
  {Kundu}(2015)}]{Chakraborty:2015raa}%
  \BibitemOpen
  \bibfield  {author} {\bibinfo {author} {\bibfnamefont {I.}~\bibnamefont
  {Chakraborty}}\ and\ \bibinfo {author} {\bibfnamefont {A.}~\bibnamefont
  {Kundu}},\ }\href {\doibase 10.1103/PhysRevD.92.095023} {\bibfield  {journal}
  {\bibinfo  {journal} {Phys. Rev.}\ }\textbf {\bibinfo {volume} {D92}},\
  \bibinfo {pages} {095023} (\bibinfo {year} {2015})},\ \Eprint
  {http://arxiv.org/abs/1508.00702} {arXiv:1508.00702 [hep-ph]} \BibitemShut
  {NoStop}%
\bibitem [{\citenamefont {Chakrabarty}\ \emph {et~al.}(2014)\citenamefont
  {Chakrabarty}, \citenamefont {Dey},\ and\ \citenamefont
  {Mukhopadhyaya}}]{Chakrabarty:2014aya}%
  \BibitemOpen
  \bibfield  {author} {\bibinfo {author} {\bibfnamefont {N.}~\bibnamefont
  {Chakrabarty}}, \bibinfo {author} {\bibfnamefont {U.~K.}\ \bibnamefont
  {Dey}}, \ and\ \bibinfo {author} {\bibfnamefont {B.}~\bibnamefont
  {Mukhopadhyaya}},\ }\href {\doibase 10.1007/JHEP12(2014)166} {\bibfield
  {journal} {\bibinfo  {journal} {JHEP}\ }\textbf {\bibinfo {volume} {12}},\
  \bibinfo {pages} {166} (\bibinfo {year} {2014})},\ \Eprint
  {http://arxiv.org/abs/1407.2145} {arXiv:1407.2145 [hep-ph]} \BibitemShut
  {NoStop}%
\bibitem [{\citenamefont {Baak}\ \emph {et~al.}(2014)\citenamefont {Baak},
  \citenamefont {Cúth}, \citenamefont {Haller}, \citenamefont {Hoecker},
  \citenamefont {Kogler}, \citenamefont {Mönig}, \citenamefont {Schott},\ and\
  \citenamefont {Stelzer}}]{Baak:2014ora}%
  \BibitemOpen
  \bibfield  {author} {\bibinfo {author} {\bibfnamefont {M.}~\bibnamefont
  {Baak}}, \bibinfo {author} {\bibfnamefont {J.}~\bibnamefont {Cúth}},
  \bibinfo {author} {\bibfnamefont {J.}~\bibnamefont {Haller}}, \bibinfo
  {author} {\bibfnamefont {A.}~\bibnamefont {Hoecker}}, \bibinfo {author}
  {\bibfnamefont {R.}~\bibnamefont {Kogler}}, \bibinfo {author} {\bibfnamefont
  {K.}~\bibnamefont {Mönig}}, \bibinfo {author} {\bibfnamefont
  {M.}~\bibnamefont {Schott}}, \ and\ \bibinfo {author} {\bibfnamefont
  {J.}~\bibnamefont {Stelzer}} (\bibinfo {collaboration} {Gfitter Group}),\
  }\href {\doibase 10.1140/epjc/s10052-014-3046-5} {\bibfield  {journal}
  {\bibinfo  {journal} {Eur. Phys. J.}\ }\textbf {\bibinfo {volume} {C74}},\
  \bibinfo {pages} {3046} (\bibinfo {year} {2014})},\ \Eprint
  {http://arxiv.org/abs/1407.3792} {arXiv:1407.3792 [hep-ph]} \BibitemShut
  {NoStop}%
\bibitem [{\citenamefont {Misiak}\ and\ \citenamefont
  {Steinhauser}(2017)}]{Misiak:2017bgg}%
  \BibitemOpen
  \bibfield  {author} {\bibinfo {author} {\bibfnamefont {M.}~\bibnamefont
  {Misiak}}\ and\ \bibinfo {author} {\bibfnamefont {M.}~\bibnamefont
  {Steinhauser}},\ }\href {\doibase 10.1140/epjc/s10052-017-4776-y} {\bibfield
  {journal} {\bibinfo  {journal} {Eur. Phys. J.}\ }\textbf {\bibinfo {volume}
  {C77}},\ \bibinfo {pages} {201} (\bibinfo {year} {2017})},\ \Eprint
  {http://arxiv.org/abs/1702.04571} {arXiv:1702.04571 [hep-ph]} \BibitemShut
  {NoStop}%
\bibitem [{\citenamefont {Misiak}\ \emph {et~al.}(2015)\citenamefont {Misiak}
  \emph {et~al.}}]{Misiak:2015xwa}%
  \BibitemOpen
  \bibfield  {author} {\bibinfo {author} {\bibfnamefont {M.}~\bibnamefont
  {Misiak}} \emph {et~al.},\ }\href {\doibase 10.1103/PhysRevLett.114.221801}
  {\bibfield  {journal} {\bibinfo  {journal} {Phys. Rev. Lett.}\ }\textbf
  {\bibinfo {volume} {114}},\ \bibinfo {pages} {221801} (\bibinfo {year}
  {2015})},\ \Eprint {http://arxiv.org/abs/1503.01789} {arXiv:1503.01789
  [hep-ph]} \BibitemShut {NoStop}%
\bibitem [{CMS(2015)}]{CMS-PAS-HIG-15-002}%
  \BibitemOpen
  \href {https://cds.cern.ch/record/2053103} {\emph {\bibinfo {title}
  {{Measurements of the Higgs boson production and decay rates and constraints
  on its couplings from a combined ATLAS and CMS analysis of the LHC pp
  collision data at $\sqrt{s}$ = 7 and 8 TeV}}}},\ \bibinfo {type} {Tech.
  Rep.}\ \bibinfo {number} {CMS-PAS-HIG-15-002. ATLAS-CONF-2015-044}\ (\bibinfo
   {institution} {CERN},\ \bibinfo {address} {Geneva},\ \bibinfo {year}
  {2015})\BibitemShut {NoStop}%
\bibitem [{\citenamefont {Keus}\ \emph {et~al.}(2014)\citenamefont {Keus},
  \citenamefont {King},\ and\ \citenamefont {Moretti}}]{Keus:2013hya}%
  \BibitemOpen
  \bibfield  {author} {\bibinfo {author} {\bibfnamefont {V.}~\bibnamefont
  {Keus}}, \bibinfo {author} {\bibfnamefont {S.~F.}\ \bibnamefont {King}}, \
  and\ \bibinfo {author} {\bibfnamefont {S.}~\bibnamefont {Moretti}},\ }\href
  {\doibase 10.1007/JHEP01(2014)052} {\bibfield  {journal} {\bibinfo  {journal}
  {JHEP}\ }\textbf {\bibinfo {volume} {01}},\ \bibinfo {pages} {052} (\bibinfo
  {year} {2014})},\ \Eprint {http://arxiv.org/abs/1310.8253} {arXiv:1310.8253
  [hep-ph]} \BibitemShut {NoStop}%
\end{thebibliography}%
  

\end{document}